\documentclass[journal]{IEEEtran}
\usepackage{amsmath,amsfonts}
\usepackage{graphicx}
\usepackage{amssymb}
\usepackage{siunitx}
\usepackage{cite}
\usepackage{mathtools}
\usepackage{bbold}
\usepackage{steinmetz}
\usepackage{stmaryrd}
\usepackage{bbm, dsfont}
\usepackage{amsthm}
\usepackage{verbatim}
\usepackage[version=4]{mhchem}
\usepackage[latin1]{inputenc}
\usepackage{tikz}
\usepackage{textcomp}
\usepackage{dirtytalk}
\usepackage{enumerate}
\usepackage{color}
\usepackage{pgfplots}
\usetikzlibrary{calc}
\usepackage{mathrsfs}
\usepackage{upgreek}
\usepackage{steinmetz}
\usepackage[outdir=./]{epstopdf}
\usepackage{lipsum}
\usepackage{amsbsy}
\usepackage{bm}
\usetikzlibrary{positioning}% To get more advances positioning options
\usetikzlibrary{shapes,arrows}
\usepackage{multirow}

\pgfplotsset{
	every tick label/.append style={scale=1},
	every axis/.append style={
	}
}

\pgfplotsset{
	grid style = {
		dash pattern = on 0.05mm off 1mm,
		line cap = round,
		black,
		line width = 0.5pt
	}
}
\usetikzlibrary{shapes.multipart,intersections}

\newcommand{\tx}{\textnormal}

\newsavebox\myboxA
\newsavebox\myboxB
\newlength\mylenA

\newcommand*\xoverline[2][0.75]{%
    \sbox{\myboxA}{$\m@th#2$}%
    \setbox\myboxB\null% Phantom box
    \ht\myboxB=\ht\myboxA%
    \dp\myboxB=\dp\myboxA%
    \wd\myboxB=#1\wd\myboxA% Scale phantom
    \sbox\myboxB{$\m@th\overline{\copy\myboxB}$}%  Overlined phantom
    \setlength\mylenA{\the\wd\myboxA}%   calc width diff
    \addtolength\mylenA{-\the\wd\myboxB}%
    \ifdim\wd\myboxB<\wd\myboxA%
      \rlap{\hskip 0.5\mylenA\usebox\myboxB}{\usebox\myboxA}%
    \else
        \hskip -0.5\mylenA\rlap{\usebox\myboxA}{\hskip 0.5\mylenA\usebox\myboxB}%
    \fi}
\makeatother

\usepackage{subfigure} % For subfigures
\usepackage{graphicx}

\begin{document}
\bstctlcite{IEEEexample:BSTcontrol}

\title{Optimal Thermal Management, Charging, and Eco-driving of Battery Electric Vehicles}

%\title{Eco-driving and Optimal Battery Thermal Management of Battery Electric Vehicles Driving in a Cold Climate}

\author{Ahad Hamednia, Nikolce Murgovski, Jonas Fredriksson, Jimmy Forsman, Mitra Pourabdollah, and Viktor Larsson  % <-this % stops a space
\thanks{A. Hamednia, N. Murgovski and Jonas Fredriksson are with the Department
of Electrical Engineering, Chalmers University of Technology,
Gothenburg 412 96, Sweden (e-mail: ahad.hamednia@chalmers.se).}% <-this % stops a space
\thanks{J. Forsman, M. Pourabdollah, and V. Larsson are with the department of Vehicle Energy and Motion Control, Volvo Car Corporation, Gothenburg 405 31, Sweden.}
}
\maketitle

\begin{abstract}
This paper addresses optimal battery thermal management (BTM), charging, and eco-driving of a battery electric vehicle (BEV) with the goal of improving its grid-to-meter energy efficiency. Thus, an optimisation problem is formulated, aiming at finding the optimal trade-off between trip time and charging cost. The formulated problem is then transformed into a hybrid dynamical system, where the dynamics in driving and charging modes are modeled with different functions and with different state and control vectors. Moreover, to improve computational efficiency, we propose modelling the driving dynamics in a spatial domain, where decisions are made along the traveled distance. Charging dynamics are modeled in a temporal domain, where decisions are made along a normalized charging time. The actual charging time is modeled as a scalar variable that is optimized simultaneously with the optimal state and control trajectories, for both charging and driving modes. The performance of the proposed algorithm is assessed over a road with a hilly terrain, where two charging possibilities are considered along the driving route. According to the results, trip time including driving and charging times, is reduced by \SI{44}{\%}, compared to a case without battery active heating/cooling.

%the proposed algorithm is applied to the vehicle.
%with one intermediate and a terminal charging. According to the simulation results, the battery temperature rate is increased by high-voltage coolant heater (HVCH) prior to the intermediate and terminal charging. This brings the battery temperature to (or closer to) its desired range with high charging power capability, which leads to a short charging time. Also, during the charging HVCH and heating,
\end{abstract}


\IEEEpeerreviewmaketitle

\section{Introduction}\label{sec:intro}

% \red{Among all parts of the electric powertrain, electric battery is the key component because:
% \begin{itemize}
%     \item The battery is the most expensive electrical component of the powertrain.
%     \item The vehicle's electric range relies almost entirely on the battery.
%     \item The battery is a heavy component, i.e. it can weigh up to about $\SI{500}{kg}$, hence it has a large thermal mass.
%     \item It is expensive to actively heat/cool the battery.
%     \item The time constant of battery temperature rate while the heating/cooling can be very long.
% \end{itemize}
% Thus, it is crucial to model the battery to study its electric and thermal behaviors.}
% \vspace{1cm}
% Electric battery is the key component among all parts of the electric powertrain, because: (i) the battery is the most expensive electrical component of the powertrain; (ii) the vehicle's electric range relies almost entirely on the battery; (iii) the battery is a heavy component, i.e. it can weigh up to about $\SI{500}{kg}$, hence it has a large thermal mass; (iv) the time constant of battery temperature rate while the heating/cooling can be very long.

\IEEEPARstart{E}{lectric} vehicles (EVs) have recently emerged as a leading technology to fulfill the increasingly stringent legislation against greenhouse gas emissions, and to counteract combustion engine vehicles' associated drawbacks, such as air pollution, climate change, high operating and maintenance costs, and recent high oil price~\cite{jaguemont15}. These issues as well as recent advances in battery technology propel vehicle manufacturers towards electromobility, aiming at developing more sustainable vehicles~\cite{yi18}. However, electromobility confronts several issues hindering the widespread use of EVs. Among them, the limited electric range of EVs is a majorn concern, which emphasizes the significance of reducing total energy consumption~\cite{oecditf19}. Also, lithium-ion (Li-ion) batteries, as a commonly used choice in the market, are highly temperature sensitive, i.e. Li-ions have reduced performance at subzero and very high temperatures, i.e. $\SI{45}{^\circ C}-\SI{60}{^\circ C}$~\cite{zhu18}. Thus, developing a suitable battery thermal management (BTM) for the electric powertrain is another hindrance to ponder on.
%Among them, ever growing vehicular energy consumption and battery thermal management (BTM) are recognized as major issues for consideration~\cite{amini19}. 

One promising way to reduce the EVs' total energy consumption is by improving \textit{grid-to-meter} efficiency, referred to as the conversion of electrical energy drawn from the electrical grid into kinetic and potential energies required for the vehicle's movement, and accompanied losses. To do so, a suggested way in the literature is to follow the principles of \textit{eco-driving}, ~\cite{kamal11}. Eco-driving can be achieved by optimising velocity profile of the vehicle given the road conditions and traffic situation. In case of driving in a hilly terrain, the optimal speed has a varying behaviour, where the vehicle typically decelerates when  climbing uphill, and accelerates when rolling downhill. This reduces non-recuperable energy waste at the braking pads, compared to driving with a constant speed~\cite{barkenbus2010eco}. To obtain an eco-driving velocity profile over complex road topographies, model-based optimal control strategies are employed to optimally coordinate energy use, see e.g.,~\cite{murgovski16,johannesson15a,hovgard18,hamedniaitsc21}. Dynamic programming (DP)~\cite{bellman57} is a widely used approach in eco-driving applications~\cite{heppeler16a,ding18,maamria19} due to its capability of solving mixed-integer, non-convex, and nonlinear optimisation problems. However, the main drawback of DP is \textit{curse of dimensionality}, i.e. computational time increases exponentially with the dimension of the optimal control problem (OCP). For high-dimensional OCPs, it is possible to reduce computational complexity by adjoining system state dynamics to the cost function and neglecting the state constraints~\cite{keulen11}, as suggested by Pontryagin's Maximum Principle (PMP)~\cite{pontryagin62,hartl95}. In~\cite{shen20} PMP is used for solving an OCP describing the driving mission with incorporated real-world considerations, e.g. speed limits and safety. A PMP-DP method is devised for optimal speed control and energy management of hybrid electric vehicles (HEVs)~\cite{uebel17}. Nonlinear programming (NLP) is another approach employed to investigate the eco-driving problem and trip time under various traffic situations~\cite{hamedniaitsc20}. In this context, multi-level or bi-level control architectures are also proposed within a model predictive control (MPC) framework in~\cite{turri16,chen18,uebel19,hamednia21}, aiming at improving computational efficiency. Thus, different tasks, for e.g. gear optimisation or disturbance rejection, are assigned to distinct layers according to time constants, updating frequency, horizon length, and sampling interval. Also, eco-driving can be used within the MPC framework for heavy-duty platooning, as shown in~\cite{sharma20}. Despite extensive contributions of eco-driving, this portion of conducted research lacks the influence from a limited range capability.
%Despite extensive contributions of eco-driving, this portion of conducted research lacks studying the BTM's impact on improving the grid-to-meter energy efficiency.

Another challenge impeding the deployment of EVs is the development of a battery management system that satisfies strict requirements on durability, performance, and safety. At high battery temperatures, the battery performance is deteriorated due to overexposure to heat, i.e. excessive battery temperatures can create sparks, flames, bulge and bubbles, and lead to battery corrosion and even explosion~\cite{hannan2018}. Also, at sub-zero temperatures, the electro-chemical process is severely slowed due to an increase in internal impedance of the battery cell. This leads to a drastic loss of the cell's available power and energy~\cite{jeffs19}. Thus, it is essential to develop an adequate BTM system, especially in places where temperature drops to sub-zero values for a considerable period of time in a year~\cite{jaguemont15}. Within the BTM system, several components, e.g. HVAC, HVCH, and heat pump, are utilised for controlling the battery pack's temperature. As these components draw power from the battery, it is pivotal to incorporate the BTM in optimising the EV's grid-to-meter energy efficiency. This leads to an increased awareness on total demanded power of the vehicle to achieve a more energy efficient drive~\cite{amini19,chen20}. Thus, various research efforts have been carried out on developing a BTM by optimal control strategies. In~\cite{jaguemont16} a DP algorithm is applied for the BTM of an electrified vehicle parked outside at low temperatures, and unplugged from the electrical grid. The algorithm's objective is to maximise the available energy in the battery pack when the vehicle is about to restart, and minimise the cell degradation stemming from low temperatures. Also, PMP is used in~\cite{bauer14} to find an optimal compromise between battery life expectancy and energy cost. Furthermore, several BTM strategies are developed within an MPC framework for achieving energy savings due to optimal cooling/heating~\cite{zhu18real,zhu18,zhu2018robust,lopez16,lopez17}. Moreover, the BTM is addressed in~\cite{chen2020joint,amini19}, where the vehicle speed profile is known a priori~\cite{chen2020joint}, or future speed prediction is included into the energy efficiency improvement OCP~\cite{amini19}. Although a vast portion of research has been carried out on BTMs, to the best of our knowledge, the optimal coordination of eco-driving and BTM for a BEV driving in a cold environment has not been explored, especially for long driving missions where optimal trade-off has to be made between travel time, energy efficiency and charging cost. 
%focusing on enhancing the grid-to-meter efficiency has not been explored. This can be highly crucial especially when the vehicle is driving in a cold environment, and with a cold initial battery temperature.

This paper considers a BEV with a cold initial battery temperature and where the ambient temperature is low throughout the vehicle's driving mission. %Thus, the drastic loss of the battery cell's provided power and energy due to the cold battery temperature is concerned in this study. 
We consider driving on a long road with a hilly terrain, where the vehicle's electric range is not sufficient to reach the destination. This necessitates assuming multiple intermediate (and terminal) charging options along the driving route. In addition to the battery temperature, the maximum available cell power is also dependent on SoC, i.e. as SoC increases, charging power capability decreases and discharging power capability increases. Furthermore, constraints on state variables and control inputs as well as governing dynamics describing the vehicle's behaviour in driving and charging modes, generally differ. % with the ones in other mode. 
If not formulated with care, the optimal control problem for optimizing eco-driving, charging, and BTM may suffer several computational issues. % hindering the investigation of the BEV's eco-driving and optimal BTM with respect to trip time when capturing both driving and charging modes. 
These include: (1) the time instants that belong to the charging and driving modes are not known prior to the vehicle's mission. Thus, there is no explicit clue about using the state variables, control inputs, constraints, and governing dynamics of each mode; (2) the vehicle longitudinal dynamics is nonlinear with respect to trip time, as the aerodynamic drag has quadratic dependency to the vehicle speed. Also, the road slope
can be an arbitrary nonlinear function of distance. Furthermore, the speed limits can have abrupt changes for some segments of the road. Accordingly, the speed limits may be non-smooth and non-differentiable functions of travel distance. %Thus, it is crucial to overcome these drawbacks when formulating the optimisation problem.

%In line with the afore-mentioned goals, the main contributions of the present research are summarized as follows:
%To overcome above-mentioned drawbacks, and to achieve optimal BTM, eco-driving, and charging of the BEV, the main contributions of the present research are summarized as follows:
%\begin{itemize}
To overcome above-mentioned computational drawbacks and achieve optimal BTM, charging, and eco-driving, under the low power availability of the battery cells at cold temperatures, we propose an optimisation problem formulated as a hybrid dynamical system. Within the problem formulation, the dynamics in driving and charging modes are modeled with different state and control vectors, and with different functions. The driving dynamics are modeled in a spatial domain, i.e. decisions are made along the traveled distance. Also, charging dynamics are modeled in a temporal domain, i.e. decisions are made along a normalized charging time. The actual charging time is optimized together with the optimal state and control trajectories, for both charging and driving modes. Within the problem formulation, multiple intermediate (and terminal) charging possibilities are included along the route, to increase scalability and feasibility of the developed algorithm in expressing more realistic driving situations.  
    %\item To enhance the awareness on vehicular complete energy analysis and strive for improving grid-to-meter energy efficiency, an optimisation problem is primarily formulated in terms of trip time to jointly capture eco-driving and optimal BTM, subject to road and traffic information. In the problem formulation, battery discharge and charge power limits for a given combination of battery temperature and state of charge (SoC) are incorporated. These are essential in finding an optimal compromise among electrical energy cost during driving, charging energy cost, and trip time.
    %\item To alleviate the computational challenges, the optimisation problem is transformed into a hybrid dynamical system, where each mode is described with a distinct independent variable. %More details are given in Section~\ref{sec:method}.
%\end{itemize}

The rest of the paper is organized as follows. Section~\ref{sec:model} addresses the overall vehicle model including longitudinal dynamics and multi-domain powertrain structure. Section~\ref{sec:pf} corresponds to the problem formulation in a temporal domain. Section~\ref{sec:method} proposes the hybrid dynamical system with the goal of alleviating computational drawbacks. In Section~\ref{sec:res} simulation results are presented. Finally, Section~\ref{sec:con} concludes the paper and outlines the possible future research directions.

% \vspace{5cm}
% Among all parts of the electric powertrain, electric battery is the key component because:
% \begin{itemize}
%     \item The battery is the most expensive electrical component of the powertrain.
%     \item The vehicle's electric range relies almost entirely on the battery.
%     \item The battery is a heavy component, i.e. it can weigh up to about $\SI{500}{kg}$, hence it has a large thermal mass.
%     \item It is expensive to actively heat/cool the battery.
%     \item The time constant of battery temperature rate while the heating/cooling can be very long.
% \end{itemize}
% Thus, it is crucial to model the battery to study its electric and thermal behaviors.
% \vspace{1cm}
% Electric battery is the key component among all parts of the electric powertrain, because: (i) the battery is the most expensive electrical component of the powertrain; (ii) the vehicle's electric range relies almost entirely on the battery; (iii) the battery is a heavy component, i.e. it can weigh up to about $\SI{500}{kg}$, hence it has a large thermal mass; (iv) the time constant of battery temperature rate while the heating/cooling can be very long.

\begin{figure}[t!]
 \centering
 \includegraphics[width=.9\linewidth]{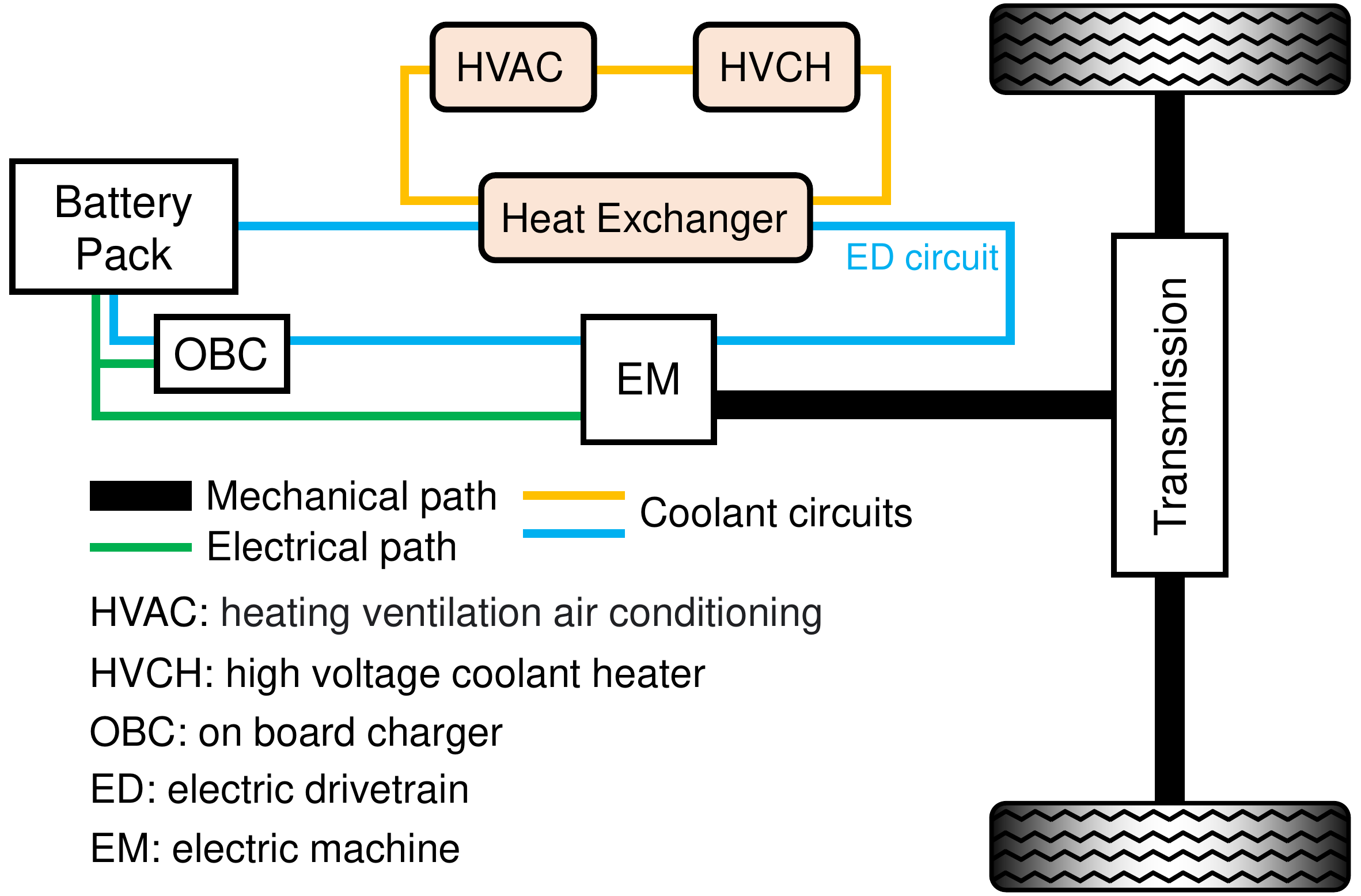}%\vspace{-.3cm}
  \caption{\footnotesize Schematic diagram of the studied electric powertrain, which consists of a  battery, an EM, a  transmission system, a thermal management system, and an on board charger. The thermal management system includes HVCH and HVAC, which are actively regulating the battery pack and cabin compartment temperatures.}
  \label{fig:powertrain}
\end{figure}

\section{Modelling}\label{sec:model}
In this section, dynamics of a BEV %treated as a lumped mass system 
is addressed. A multi-domain configuration of an electric powertrain is described, including powertrain components connecting via electrical, thermal, and mechanical paths.

\subsection{Vehicle as a Lumped Mass System}\label{subsec:lumped-mass}
According to Newton's law of motion, longitudinal dynamics of the vehicle is described by
\begin{align}
    \dot v(t)=a_\tx{t}(t)-a_\tx{air}(v(t))-a_\alpha(s(t)),\label{eq:long_dyn_t}
\end{align}
where $v$ is the vehicle's speed, $a_\tx{t}$ is traction acceleration at the wheel side of the vehicle, and $a_\tx{air}$ and $a_\alpha$ are the accelerations associated with air drag and roll and gravitational loads, respectively, as
\begin{align}
    &a_\tx{air}(v(t))=\frac{\rho_\tx{a}c_\tx{d}A_\tx{f}v^2(t)}{2m}, \label{eq:aero}\\
    &a_\alpha(s(t))=g\Big(\sin\big(\alpha(s(t)))+c_\tx{r}\cos(\alpha(s(t))\big)\Big), \label{eq:rol_pot}
\end{align}
where $\rho_\tx{a}$ is air density, $c_\tx{d}$ is aerodynamic drag coefficient, $A_\tx{f}$ is the vehicle's frontal area, $m$ is the vehicle's total lumped mass, $g$ is gravitational acceleration, $c_\tx{r}$ is rolling resistance coefficient, and $\alpha$ is road gradient. %It is possible to directly extract the road gradient from a standard global positioning system (GPS). 
%Note that the explicit dependence of travel distance and vehicle speed on trip time in \eqref{eq:long_dyn_t}-\eqref{eq:rol_pot} is not shown for brevity, although they are both functions of trip time.

The vehicle's travelled distance, $s$, is given by integrating the vehicle speed:
\begin{align}
    s(t) = \int_{0}^tv(x)dx  \ \Rightarrow \ \dot s(t)=v(t),\label{eq:time_dyn_t}
\end{align} 
where $t$ is trip time.

\begin{figure}[t!]
\centering
\includegraphics[width=0.4\linewidth]{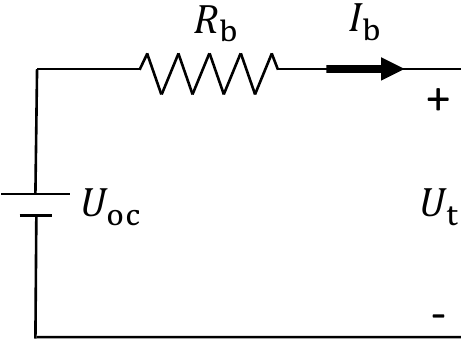}
\caption{\footnotesize Equivalent circuit of the battery pack, where $U_\tx{oc}$ is open circuit voltage, $R_\tx{b}$ is internal resistance, $I_\tx{b}$ is battery discharge current and $U_\tx{t}$ is battery terminal voltage}
\label{fig:equ_circuit}
\end{figure}

\subsection{Multi-domain Powertrain Structure}\label{subsec:powertrain}
%The generation of traction acceleration depends on the vehicle's powertrain. 
Fig.~\ref{fig:powertrain} depicts the schematic diagram of the studied electric powertrain. The powertrain consists of an electric machine (EM) as an actuator, a transmission system and a battery for energy supply or storage. Apart from the propulsion components the powertrain also consists of a thermal management system, and an on board charger (OBC). As it is demonstrated in Fig.~\ref{fig:powertrain}, the electric power flow through an electrical path is bidirectional depending on operating mode of the EM. Thus, the battery receives energy from the EM in generating mode, or delivers energy to the EM in motoring mode. HVAC and HVCH are the components used for the thermal management of cabin compartment and battery pack, i.e. HVCH and HVAC are mainly used for heating and cooling, respectively. The OBC is a device that is employed for regulating the flow of electricity from the electrical grid to the battery, monitoring the charging rate and for protection purposes. Note that the OBC is assumed to be ideal in this paper.

\subsubsection{Electrical Domain}\label{subsubsec:el_dom}
The battery is modelled using an equivalent circuit shown in Fig.~\ref{fig:equ_circuit}. The circuit includes a voltage source $U_\tx{oc}$ and an internal resistance $R_\tx{b}$, which are mainly influenced by SoC and battery temperature, respectively. The internal resistance is generally proportional to the inverse of battery temperature~\cite{zhu18}. Thus, by increasing the battery temperature, the ions inside the battery cells gain more energy, which leads to a reduction in the encountered resistance against the ions' displacement. Note that there is a slight mismatch between the internal resistance while charging and discharging, which is neglected in this work. Open-circuit voltage is commonly a nonlinear monotonically increasing function of SoC, which is usually derived via offline experiments at different battery aging stages and ambient temperatures. The change of SoC is given by
\begin{align}
    \dot{\tx{soc}}(t)=-\frac{P_\tx{b}(t)}{C_\tx{b}U_\tx{oc}(\tx{soc}(t))},
    \label{eq:soc_dyn_t}
\end{align}
where $P_\tx{b}$ is battery power including internal resistive losses, and $C_\tx{b}$ is maximum capacity of the battery. $P_\tx{b}$ is positive when discharging, and is negative while charging.

\begin{figure*}[t!]
 \centering
 \includegraphics[width=.95\linewidth]{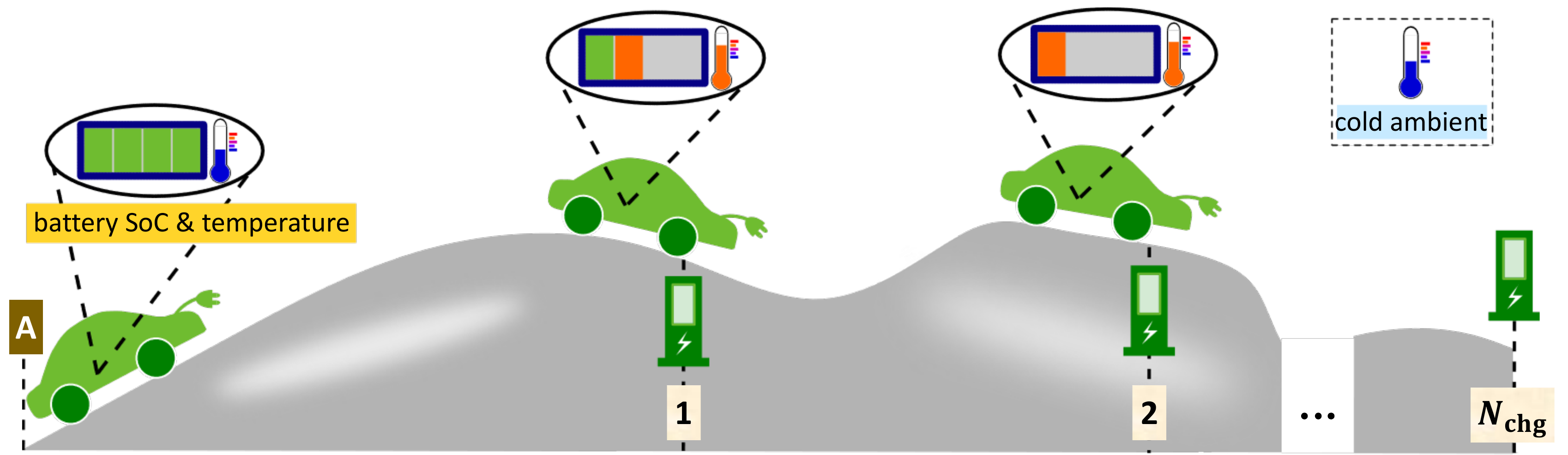}
  \caption{\footnotesize Studied scenario; a BEV is driving in a hilly terrain, where the vehicle starts its mission from point A with a fully charged battery with sub-zero ambient and initial battery temperatures. The indices $1$, $2$, and $N_\tx{chg}$ denote to the intermediate (an terminal) charging stations.}
  \label{fig:scenario}
\end{figure*}
% \begin{figure*}[ht]
%   \centering
%  \subfigure[\footnotesize]{
%  \includegraphics[width=0.175\linewidth]{Fig/equ_circuit.pdf}
%  \label{fig:equ_circuit}
%  }\hspace{1.25cm}
% %   \centering
%  \subfigure[\footnotesize]{
%  \includegraphics[width=0.275\linewidth]{Fig/Rb_Tb.pdf}
%  \label{fig:Rb}
%  }\hspace{1.25cm}
% %  \centering 
%  \subfigure[\footnotesize]{
%  \includegraphics[width=0.275\linewidth]{Fig/Uoc_soc_rot.pdf}
%  \label{fig:Uoc}
% }
% \caption{\footnotesize Battery electric model. (a) equivalent circuit of the battery pack. (b) normalised battery internal resistance. (c) normalised battery open-circuit voltage.}
% \label{fig:elcir}
% \end{figure*}

\subsubsection{Thermal Domain}\label{subsubsec:th_dom}
An energy balance is used to describe the battery pack's dynamics. Following the fundamental thermodynamic principle, a lumped-parameter thermal model describing the dynamical variations of the battery pack's temperature is given by
\begin{align}
\begin{split}
&\dot{T_\tx{b}}(t)=\frac{1}{c_\tx{p}m_\tx{b}}\big(Q^\tx{gen}_\tx{pass}(\cdot)+Q_\tx{act}(P_\tx{hvch}^\tx{b}(t),P_\tx{hvac}^\tx{b}(t))\\
&\hspace{3.5cm}+Q_\tx{exh}(T_\tx{b}(t),T_\tx{amb}(t),v(t))\big),
\end{split}
\end{align}
where $c_\tx{p}$ is specific heat capacity of the battery pack, $m_\tx{b}$ is total battery mass, the symbol $\cdot$ is a  compact notation for a function of multiple variables, $Q^\tx{gen}_\tx{pass}$ is the rate of generated heat by sources that passively affect the battery temperature, $Q_\tx{act}$ is the heat rate due to components that can actively adjust the battery pack temperature, $P_\tx{hvch}^\tx{b}$ and $P_\tx{hvac}^\tx{b}$ are HVCH and HVAC powers, respectively, and $Q_\tx{exh}$ is the heat exchange rate among the battery pack, ambient air and/or the chassis of the vehicle.

The passive generated heat includes: 1) irreversible ohmic Joule heat induced by the battery internal resistive losses; and 2) heat generated by electric drivetrain (ED) power losses, $Q_\tx{ed}$. For a given pair of vehicle speed and traction acceleration; the passive generated heat rate can be written as
\begin{align}
    Q^\tx{gen}_\tx{pass}(\cdot)=R_\tx{b}(T_\tx{b}(t))\frac{P_\tx{b}^2(t)}{U_\tx{oc}^2(\tx{soc}(t))}+Q_\tx{ed}(v(t),a_\tx{t}(t))\label{eq:qrqed}.
\end{align}
%More details about modelling of $Q_\tx{ed}$ will be given later in Section.~\ref{sec:method}. 
Note that the heat losses can generally originate from two types of conductive and convective heat transfers. In this paper, the uneven conductive distribution of the battery pack temperature associated with the diffusion is overlooked to avoid increasing complexity of the thermal model. Thus, the core and crust battery pack temperatures are assumed to be identical.

The active heat rate 
\begin{align}
    Q_\tx{act}(P_\tx{hvch}^\tx{b}(t),P_\tx{hvac}^\tx{b}(t))=\eta_\tx{hvch}P_\tx{hvch}^\tx{b}(t)-\eta_\tx{hvac}P_\tx{hvac}^\tx{b}(t) \label{eq:qact}
\end{align}
corresponds to the power conversion of the HVCH and HVAC systems, respectively, with the battery pack's heating with efficiency of $\eta_\tx{hvch}$, and its cooling with efficiency of $\eta_\tx{hvac}$.
Note that cooling of cabin compartment is out of scope in this paper.

The convective heat exchange rate between the battery pack and ambient air is modelled as
\begin{align}
    Q_\tx{exh}(T_\tx{b}(t),t)=\gamma(v(t))(T_\tx{amb}(t)-T_\tx{b}(t)),\label{eq:qexh}
\end{align}
where $T_\tx{amb}$ is ambient temperature, and $\gamma>0$ is a speed dependent coefficient representing parasitic heat transfer between the battery and the ambient air, i.e. if the battery temperature is higher than the ambient temperature, heat is conveyed from the battery to the ambient air.

\subsubsection{Mechanical Domain}\label{subsubsec:prop_dom}
The EM when operated in motoring mode, provides propulsive power, which is delivered via the transmission system to the wheels through a mechanical path, see Fig.~\ref{fig:powertrain}. To do so, the EM torque and rotational speed are translated by the transmission system to traction acceleration and vehicle speed, respectively. Speed dependent bounds on EM torque are translated as limits on traction acceleration via
\begin{align}
    a_\tx{t}(t)\in[a_{\min}(v(t)),a_{\max}(v(t))].\label{eq:at_bounds_t}
\end{align}
%which are imposed by the EM torque limits.

% through the static equations
% \begin{align}
%     a_\tx{t}(t)=\frac{M(t)}{mR}, \quad v(t)=\omega(t) R, \quad R=\frac{r_\tx{w}}{r_\tx{fg}},
%     \label{eq:trn}
% \end{align}
% % where $r_\tx{w}$ is wheel radius and $r_\tx{fg}$ is final gear ratio, respectively. Speed dependent bounds on traction acceleration are
% \begin{align}
%     a_\tx{t}(t)\in[a_{\min}(v),a_{\max}(v)],\label{eq:at_bounds_t}
% \end{align}
% which are imposed by the EM torque limits.

\section{Problem Statement}\label{sec:pf}
Consider a BEV driving in a hilly terrain, as in Fig.~\ref{fig:scenario}. The trip starts from point \textbf{A} with a cold initial battery temperature and full SoC, where the  ambient temperature is also low during the vehicle's trip. As the vehicle continues its drive, the battery depletes and its temperature may increase due to the passive and/or active heating sources. The vehicle's travelled distance is greater than its range and intermediate (and terminal charging) possibilities have to be considered along the driving route.

\subsection{Bounds on Vehicle Speed, Battery Power and Grid Power}\label{subsec:vpbpgridlim}
Using available information about the road and traffic situation, the vehicle speed limits are defined as %for $i\in\mathcal{I}=\{1,2,\dots,N_\tx{chg}\}$, as
\begin{align}
v(t)\in\begin{cases}
	[v_{\min}(s(t)),v_{\max}(s(t))], & \text{$t\in \mathcal{T}_\tx{drv}$}\vspace{0.25cm}\\
    \{0\}, & \text{$t\in \mathcal{T}_\tx{chg}^{i}$}
\end{cases}\label{eq:vlim}
\end{align}
where $0<v_{\min}\leq v_{\max}$, $\mathcal{T}_\tx{drv}$ and $\mathcal{T}_\tx{chg}$ denote the sets of driving and charging time instants, respectively, ${i\in\mathcal{I}=\{1,2,\dots,N_\tx{chg}\}}$ is charger index, and $N_\tx{chg}$ is total number of charging stations along the route. 

\begin{figure*}[t!]
\centering
\subfigure[Battery discharge power limit.]{
 \includegraphics[width=.425\linewidth]{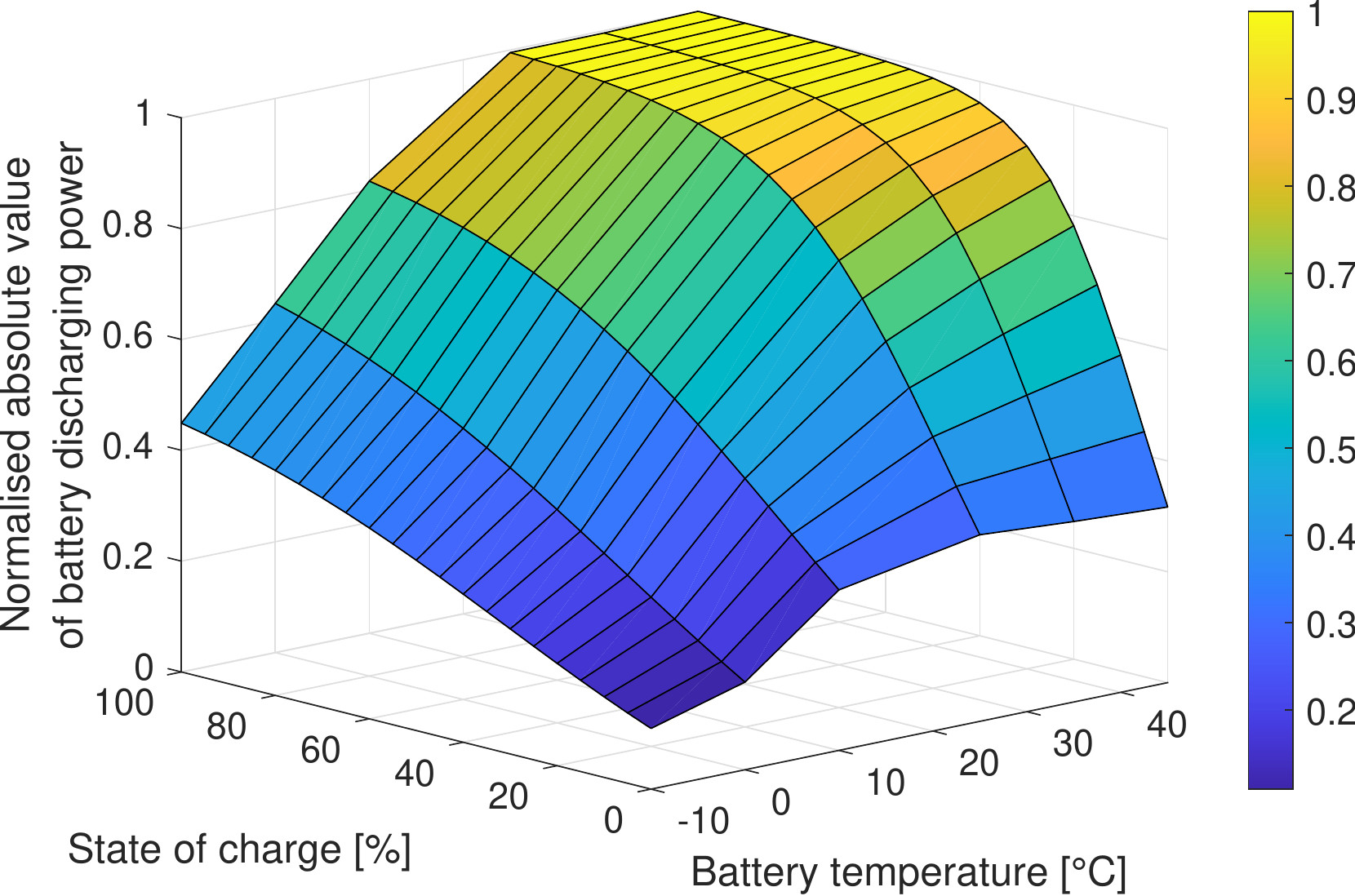}\hspace{1cm}
\label{fig:pbdchglim}
}
\subfigure[Battery charge power limit.]{

 \includegraphics[width=.425\linewidth]{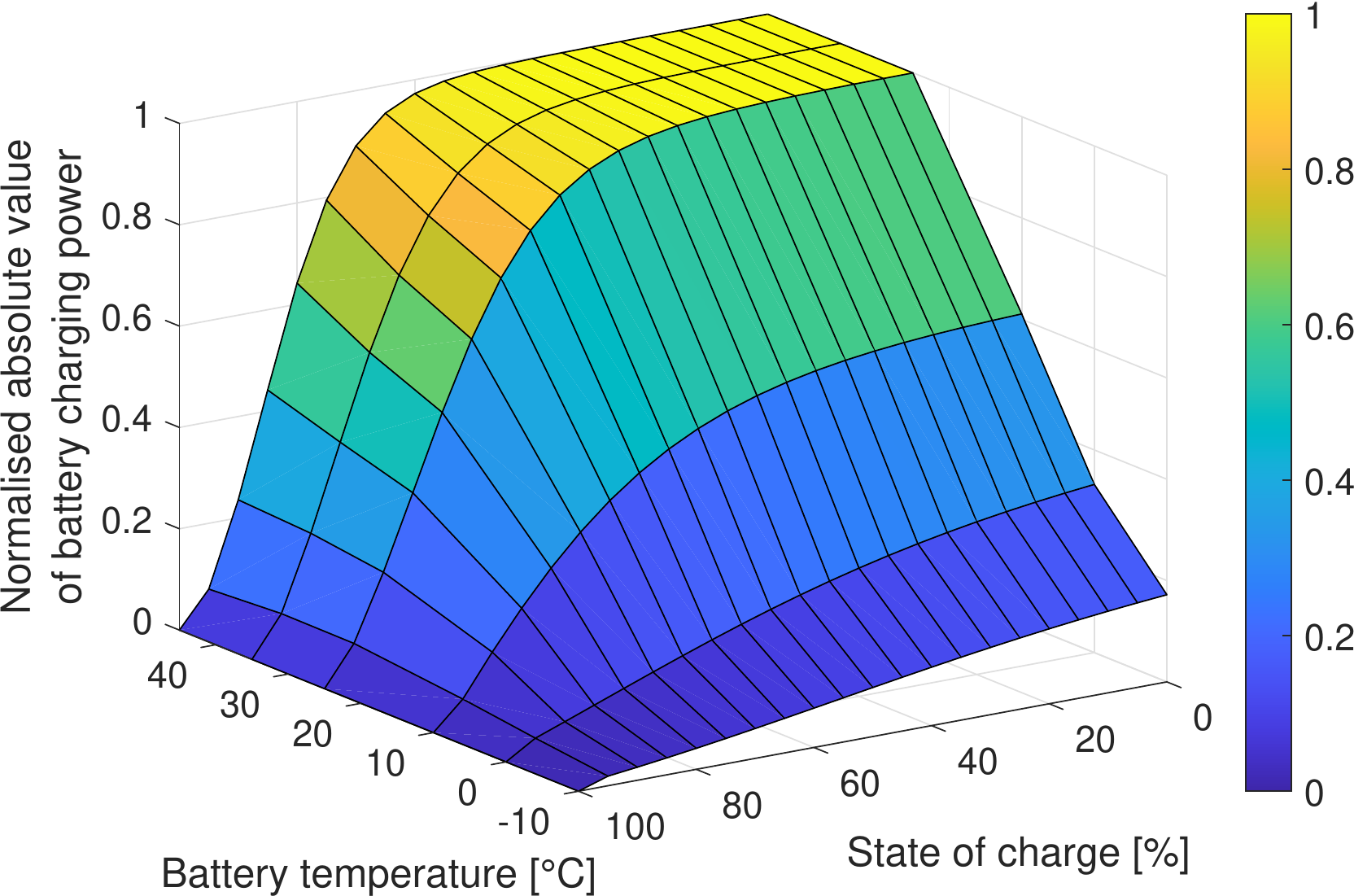}
\label{fig:pbchglim}
}
\caption{Normalised absolute value of battery charge and  discharge power limit for a given combination of battery temperature and SoC.}
\label{fig:pblim}
\end{figure*}

The speed limits include legal and dynamic speed limits that resemble realistic driving situations. New modern technologies, e.g. e-horizon systems, can provide the information about legal and dynamic speed limits and the road slope~\cite{lindgarde16}. The dynamic speed limits are enforced due to presence of e.g. intersections, ramps, junctions and  traffic lights. The legal speed limits may have abrupt changes for different segments of the driving road, where such variations can lead to computational issues that are discussed later in this Section~\ref{sec:pf} and Section~\ref{sec:method}. Note that the vehicle speed is equal to zero when the vehicle stops at the charging station. %We assume that the vehicle is equipped with a GPS and road slope, including speed limits (including dynamic) are known.

%A thorough investigation of traffic situation impact on vehicular energy efficiency and trip time has been performed as our previous work in \cite{hamedniaitsc20}.

For a given pair of battery temperature and SoC, the battery power limits corresponding to driving and charging modes for $i\in\mathcal{I}$ are given by
\begin{align}
\resizebox{0.96\hsize}{!}{$P_\tx{b}(t)\in \begin{cases}
	[P_\tx{b,chg}^{\min}(\tx{soc}(t),T_\tx{b}(t)),P_\tx{b,dchg}^{\max}(\tx{soc}(t),T_\tx{b}(t))], & \text{$t\in \mathcal{T}_\tx{drv}$}\vspace{0.25cm}\\
    [P_\tx{b,chg}^{\min}(\tx{soc}(t),T_\tx{b}(t)),0], & \text{$t\in \mathcal{T}_\tx{chg}^{i}$}
\end{cases}$}\label{eq:pbchglim} 
\end{align}
where $P_\tx{b,dchg}^{\max}>0$ and $P_\tx{b,chg}^{\min}<0$ are the battery discharge and charge power limits, respectively. It can be deduced from \eqref{eq:pbchglim} that the battery power during driving can also be negative due to regenerative braking, referred to as a mechanism that transforms the vehicle's kinetic energy into electrical energy to be stored in the battery. Note that the charging power limit may differ in driving and charging modes. Here, we assume that the same bound is applied, for simplicity, and without loss of generality.

Normalised absolute values of the battery discharge and charge power limits versus battery temperature and SoC are illustrated in Fig.~\ref{fig:pbdchglim} and Fig.~\ref{fig:pbchglim}, respectively. As shown in Fig.~\ref{fig:pbdchglim}, the battery discharge power limit is proportional to the battery temperature and SoC level. %Note that the SoC is generally the dominant variable compared to battery temperature in determining the battery discharge power limit.
Also, the charge power limit is proportional to the battery temperature and inverse of SoC level, according to Fig.~\ref{fig:pbchglim}. For the studied battery, the desirable SoC range for the discharge and charge power limits are about $\SI{25}{\%}-\SI{100}{\%}$ and $\SI{0}{\%}-\SI{60}{\%}$, respectively. Also, the battery temperature window for attaining high power availability is about $\SI{25}{^\circ C}-\SI{45}{^\circ C}$, when both charging and discharging. Thus, for a cold battery it is generally favourable to perform battery \textit{preconditioning}, referred to as heating up a cold battery prior to charging in order to charge the battery with a high power, thereby reducing the charging time.

% The electro-chemical process is severely slowed at low battery temperatures, hence leading to a great loss of power and energy. Thus, it is generally favourable to heat up the cold battery prior to charging, in order to charge the battery with a high power, and within a short time. Also, the battery capacity is deteriorated due to overexposure to heat, i.e. excessive battery temperatures can cause to battery corrosion, and create sparks, flames, bulge and bubbles~\cite{hannan2018}. Thus, it is crucial to prevent the battery temperature from avoiding its threshold.

The power $P_\tx{grid}^i$ provided by the $i$th charger is limited by
\begin{align}
P^{i}_\tx{grid}(t) \in \begin{cases}
	\{0\}, & \text{$t\in \mathcal{T}_\tx{drv}$},\vspace{0.25cm}\\
   [0,P^{i,\max}_{\tx{grid}}], & \text{$t\in \mathcal{T}_\tx{chg}^{i}$}
\end{cases}\label{eq:pgridbound}
\end{align}
where $P_\tx{grid}^{i,\max}$ is rated power of the $i$th charger. It is here assumed that grid charging power is not supplied to the vehicle during the driving mode, although the method presented later can directly be applied to the vehicles driving on an electric road, e.g. when charging lanes are installed on the road~\cite{limb18}.

\subsection{Objective Function}\label{subsec:obj}
In order to achieve an optimal compromise between trip time and charging cost, an optimisation problem is formulated with the performance function $J$, as
\begin{align}
\begin{split}
&J(\cdot)=\sum_{i=1}^{N_\tx{chg}}\Big(\int_{t\in \mathcal{T}_\tx{chg}^{i}}^{}
c^{i}_\tx{e}P^{i}_\tx{grid}(t)\tx{d}t+c^{i}_\tx{T}\max\big(0,t^{i}_\tx{chg}-T^{i}\big)\Big)\\
&\hspace{3CM}+\int_{t\in\mathcal{T}}^{}{c_\tx{t,trip}}\tx{d}t,
\end{split}    
\end{align}
where the charging cost can be expressed as energy and/or time, depending on the pricing plan of each charging station. Thus, $J$ includes:
\begin{itemize}
   %\item Electrical energy consumption during driving, as
%   \begin{align}
%       \int_{t\in \mathcal{T}_\tx{drv}}^{}{c_\tx{e,drv}P_\tx{b}(t)}\tx{d}t,\label{eq:J_edrv}
%   \end{align}
%   where $c_\tx{e,drv}$ is currency per-kilowatt-hour electricity cost while driving.
   
   \item Electrical energy supplied to the vehicle by chargers, as  
   \begin{align}
       \sum_{i=1}^{N_\tx{chg}}\int_{t\in \mathcal{T}_\tx{chg}^{i}}^{}
       c^{i}_\tx{e}P^{i}_\tx{grid}(t)\tx{d}t,\label{eq:J_echg}
   \end{align}
   where $c_\tx{e}$ is currency per-kilowatt-hour charging electrical energy cost.
   
   \item The time based cost for occupying the charging spot, as
    \begin{align}
       \sum_{i=1}^{N_\tx{chg}}c^{i}_\tx{T}\max\big(0,t^{i}_\tx{chg}-T^{i}\big),\label{eq:J_occuchg}
   \end{align}
   where $c_\tx{T}$ is currency per-minute cost due to occupying the charger for longer time than $T\geq 0$, and $t_\tx{chg}$ is a scalar variable representing charging time.
   
   \item A penalty on total trip time, as
    \begin{align}
    \int_{t\in\mathcal{T}}^{}{c_\tx{t,trip}}\tx{d}t,\label{eq:J_triptime}
   \end{align}
   where $c_\tx{t,trip}$ is the penalty factor and $\mathcal{T}=\bigcup_{i\in\mathcal{I}}^{} \mathcal{T}_\tx{chg}^{i}\bigcup \mathcal{T}_\tx{drv}$. Note that the trip time includes the charging time; thus, charging time may need to be paid twice, due to a longer trip and/or occupying the charger.
\end{itemize}

\subsection{Optimisation Problem with Respect to Trip Time}\label{subsec:pt}
For $i\in\mathcal{I}$, the optimisation problem can now be summarised, as
{\allowdisplaybreaks
\begin{subequations} \label{eq:pt}
\begin{align}
&\min_{P^\tx{b}_\tx{hvch}, P^\tx{b}_\tx{hvac}, P_\tx{b}, P_\tx{grid}^{i}, a_\tx{t}, t_\tx{chg}^{i}} J(\cdot)\\
&\text{subject to: \eqref{eq:vlim}-\eqref{eq:pgridbound} and} \nonumber\\
\begin{split}
&\dot{T_\tx{b}}(t)=\frac{1}{c_\tx{p}m_\tx{b}}\big(Q^\tx{gen}_\tx{pass}(\cdot)+Q_\tx{act}(P_\tx{hvch}^\tx{b}(t),P_\tx{hvac}^\tx{b}(t))\\
&\hspace{1.5cm}+Q_\tx{exh}(T_\tx{b}(t),T_\tx{amb}(t),v(t))\big),\quad t\in\mathcal{T}
\end{split}\label{eq:pt_tb}\\ 
% &\dot{T_\tx{b}}(t)=\frac{1}{c_\tx{p}m_\tx{b}}\big(Q^\tx{gen}_\tx{pass}(t)+Q_\tx{act}(t)+Q_\tx{exh}(t)\big),\quad t\in\mathcal{T}\label{eq:pt_tb}\\
&\dot{\tx{soc}}(t)=-\frac{P_\tx{b}(t)}{C_\tx{b}U_\tx{oc}(\tx{soc}(t))},\quad t\in\mathcal{T}\label{eq:pt_soc}\\
&\dot s(t)=v(t),\quad t\in\mathcal{T}_\tx{drv}\label{eq:pt_s}\\
&\dot v(t)=a_\tx{t}(t)-a_\tx{air}(v(t))-a_\alpha(s(t)),\quad t\in\mathcal{T}_\tx{drv}\label{eq:pt_v}\\
\begin{split}
&P^{i}_\tx{grid}(t)+P_\tx{b}(t)=R(T_\tx{b}(t))\frac{P_\tx{b}^2(t)}{U_{\tx{oc}}^2(\tx{soc}(t))}+P_\tx{prop}(v(t),a_\tx{t}(t))\\
&\hspace{.75cm}+P^\tx{b}_\tx{hvch}(t)+P^\tx{b}_\tx{hvac}(t)+P^{\tx{c}}_{\tx{hvch}}(t)+P_{\tx{aux}}(t),\quad t\in\mathcal{T} 
\end{split}\label{eq:pt_Peq}\\
&s(t) = s_\tx{chg}^{i},\quad t\in\mathcal{T}_\tx{chg}^{i} \label{eq:schgtchg}\\
&T_\tx{b}(t) \in [T_\tx{b}^{\min}(t),T_\tx{b}^{\max}(t)],\quad t\in\mathcal{T} \label{eq:pt_tbbound}\\
&\tx{soc}(t) \in [\tx{soc}_{\min}(t),\tx{soc}_{\max}(t)],\quad t\in\mathcal{T} \label{eq:pt_socbound}\\
% &v(t)\in\begin{cases}
% 	[v_{\min}(s(t)),v_{\max}(s(t))], & \text{$t\in \mathcal{T}_\tx{drv}$}\vspace{0.25cm}\\
%     \{0\}, & \text{$t\in \mathcal{T}_\tx{chg}^{i}$}
% \end{cases}\label{eq:pt_vbound}\\
&P^\tx{b}_\tx{hvch}(t) \in [0,P^{\max}_{\tx{hvch}}-P^{\tx{c}}_{\tx{hvch}}(t)],\quad t\in\mathcal{T} \label{eq:pt_pbhvchbound}\\
&P^\tx{b}_\tx{hvac}(t) \in [0,P^{\max}_{\tx{hvac}}],\quad t\in\mathcal{T} \label{eq:pt_pbhvacbound}\\
% &\resizebox{0.96\hsize}{!}{$P_\tx{b}(t)\in \begin{cases}
% 	[P_\tx{b,drv}^{\min}(T_\tx{b}(t),\tx{soc}(t)),P_\tx{b,drv}^{\max}(T_\tx{b}(t),\tx{soc}(t))], & \text{$t\in \mathcal{T}_\tx{drv}$}\vspace{0.25cm}\\
%     [P_\tx{b,chg}^{\min}(T_\tx{b}(t),\tx{soc}(t)),0), & \text{$t\in \mathcal{T}_\tx{chg}^{i}$}
% \end{cases}$}\label{eq:pt_Pbbound}\\
% &P^{i}_\tx{grid}(t) \in \begin{cases}
% 	\{0\}, & \text{$t\in \mathcal{T}_\tx{drv}$},\vspace{0.25cm}\\
%   (0,P^{i,\max}_{\tx{grid}}], & \text{$t\in \mathcal{T}_\tx{chg}^{i}$}
% \end{cases}\label{eq:pt_Pgridbound}\\
&a_\tx{t}(t)\in [a_{\min}(v(t)),a_{\max}(v(t))],\quad t\in\mathcal{T}_\tx{drv}\label{eq:pt_atbound}\\
&t_\tx{chg}^{i}\in [0,t_\tx{chg}^{\max}]\label{eq:pt_tchgbound}\\
&T_\tx{b}(0)=T_\tx{b0}, \ \tx{soc}(0)=\tx{soc}_0, \ s(0)=s_0, \ v(0)=v_0\label{eq:pt_int}\\
&T_\tx{b}(t_\tx{f})\geq T_\tx{bf}, \ \tx{soc}(t_\tx{f})\geq \tx{soc}_\tx{f}, \ s(t_\tx{f})=s_\tx{f}\label{eq:pt_final}
\end{align}
\end{subequations}}%
where $T_\tx{b0}$ and $T_\tx{bf}$ are initial and final battery temperatures, respectively, $\tx{soc}_0$ and $\tx{soc}_\tx{f}$ are initial and final SoC values, respectively, $s_0$ and $s_\tx{f}$ are initial and final travel distances, respectively, $v_0$ is initial vehicle speed, $P_\tx{prop}$ is propulsion power including the internal losses of the powertrain for a given pair of vehicle speed and traction acceleration, $P_\tx{aux}$ is given auxiliary load demand, $T_\tx{b}^{\min}$ and $T_\tx{b}^{\max}$ are the bounds on battery temperature, $\tx{soc}_{\min}$ and $\tx{soc}_{\max}$ are SoC limits, $P^{\max}_{\tx{hvch}}$ and $P^{\max}_{\tx{hvac}}$ are the maximum deliverable HVCH and HVAC power values, respectively, $s_\tx{chg}$ is the charging position that is known prior to starting the vehicle's driving mission, $t_\tx{chg}^{\max}$ is the maximum allowed charging time, and $P^{\tx{c}}_{\tx{hvch}}$ is the HVCH power demand for heating the cabin compartment. Note that the $P^{\tx{c}}_{\tx{hvch}}$ is assumed to be a function of the known ambient temperature.

\begin{figure}[t!]
 \centering
 \includegraphics[width=\linewidth]{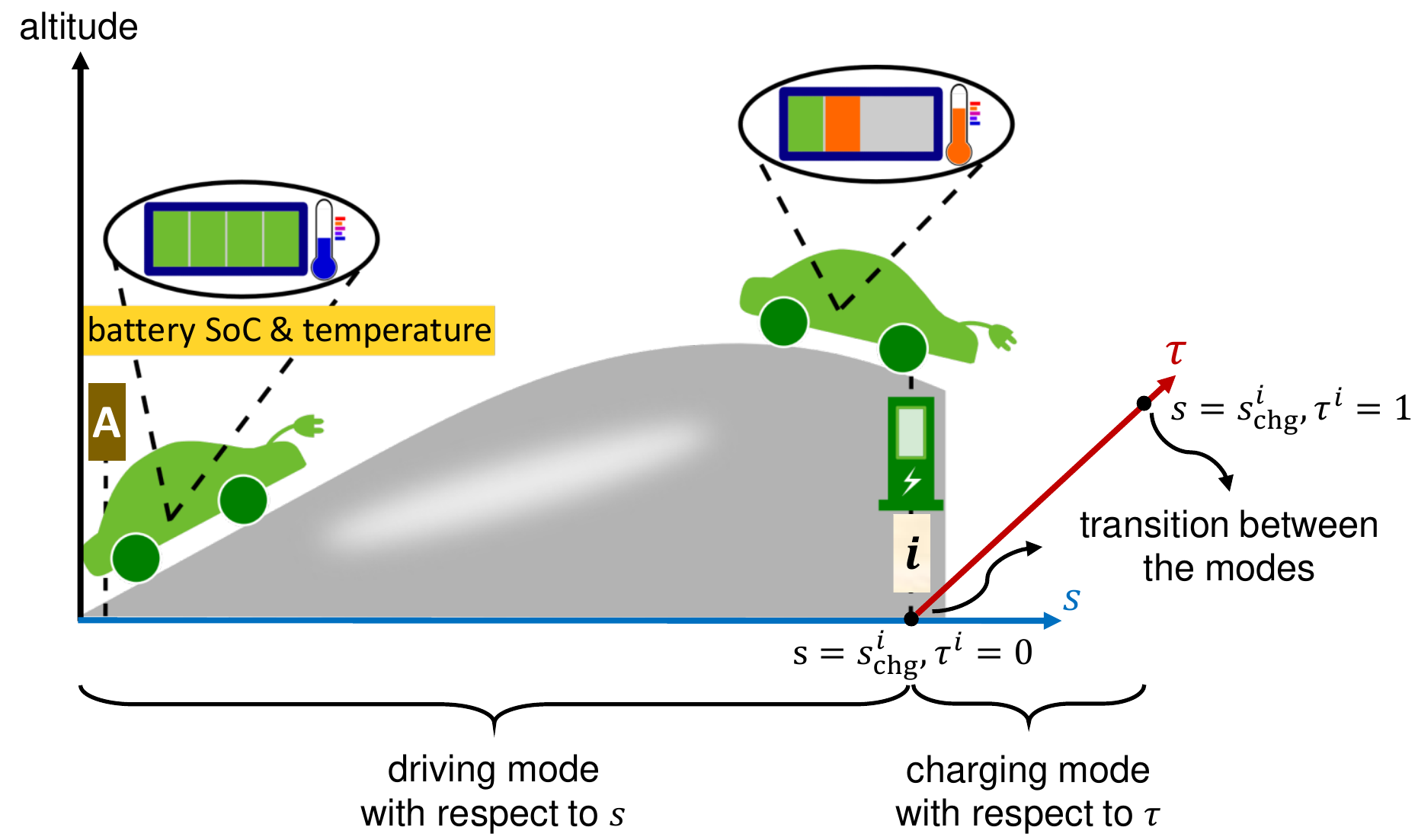}
  \caption{\footnotesize Schematic illustration of a hybrid dynamical system; driving mode, charging mode and transition between these two modes. During the driving mode decisions are taken in terms of $s$, and in charging mode the decisions are planned with respect to $\tau^i$, $i\in\mathcal{I}$.}
  \label{fig:module}
\end{figure}

The full problem \eqref{eq:pt}, including the formulations of both driving and charging modes with respect to $t$ %, is an NLP, which is extremely 
is difficult to solve due to the following reasons:
\begin{itemize}
    \item The sets including charging mode and driving mode time instants, $\mathcal{T}_\tx{chg}$ and $\mathcal{T}_\tx{drv}$, respectively, are unknown prior to the optimisation. Thus, imposing the right dynamics/values/bounds in \eqref{eq:pt} may require introducing integer variables, which would make the problem intractable.
    \item The vehicle longitudinal dynamics \eqref{eq:long_dyn_t} is nonlinear with respect to $t$, as the aerodynamic drag is quadratically dependent to vehicle speed in \eqref{eq:aero}, the road gradient can be any arbitrary nonlinear function of $t$ in \eqref{eq:rol_pot}, and the speed limits \eqref{eq:vlim} may also be non-smooth functions of $s$, i.e. the speed limits can generally change abruptly for different segments of the driving road. This may require additional integer variables, or smoothing techniques.
\end{itemize}
The aforementioned issues can severely increase computational complexity. Thus, we propose several reformulation steps in Section~\ref{sec:method} that transform the problem \eqref{eq:pt} into a hybrid dynamical system that can be solved in a minute or less on a standard computer.%, which is solved with significantly enhanced computational efficiency compared to the original problem.

\section{Hybrid Dynamical System Formulation}\label{sec:method}
In this section, the highlighted issues in Section~\ref{sec:pf} are resolved by exact reformulations of driving and charging modes:
\begin{itemize}
    \item \textbf{Driving mode}: During driving mode, $s$ is chosen as an independent variable instead of $t$, i.e. decisions are made with respect to $s$, as depicted in Fig.~\ref{fig:module}. Such transformation is valid throughout the driving mode, as the vehicle does not stop or change its direction of movement, i.e. $v>0$. Accordingly, for a certain road topography, the function $a_\alpha$ becomes a fixed trajectory covering the entire route. Also, the speed limits directly turn into position dependent limits; thus, the sudden legal speed limit change is no longer an issue. Furthermore, to remove the nonlinearity in \eqref{eq:aero}, kinetic energy of unit mass $E$ with respect to $s$ is selected instead of $v$, as
    \begin{align}
      E(s)=\frac{v^2(s)}{2}. \label{eq:v2E}
    \end{align}
    Thus, the aerodynamic drag \eqref{eq:aero} becomes a linear function of unit mass kinetic energy. Note that the decision making in the spatial coordinate is promising, since the charging positions are given. Subsequently, driving and charging distance instances are known prior to optimisation.

    \item \textbf{Charging mode}: Despite fixed position of the vehicle at the charging station, battery temperature and SoC will change during charging. Thus, the battery temperature and SoC dynamics cannot be described with respect to $s$ for the charging mode. Instead, the decisions are planned with respect to a variable $\tau^i \in [0,1]$, defined, as
    \begin{align}
    \tau^i=\frac{t}{t_\tx{chg}^{i}}, \quad t\in \mathcal{T}_\tx{chg}^{i}, i\in\mathcal{I}.\label{eq:t2tau}
    \end{align}
\end{itemize}
Following this selection of independent variables, %in order to capture the behaviour of each mode, and considering the transition between the modes, the 
problem \eqref{eq:pt} is transformed into a hybrid dynamical system, see Fig.~\ref{fig:module}. Note that state variables, control inputs and governing dynamics describing each mode may differ with those from the other mode's, which will be explained later in this Section. By repeating the combination of driving and charging modes, it is possible to investigate multiple charging scenarios along the vehicle's trip. Hereafter, the variables with subscripts or superscripts `drv' or `chg', are the previously introduced variables that now belong specifically to the driving mode or charging mode, respectively.

\subsection{Driving Mode: Dynamics and Performance Function}\label{subsec:drv_mode}
Governing dynamics during driving mode include the vehicle's longitudinal dynamics, and the dynamical variations of battery temperature and SoC. To group the state variables and control inputs belonging to driving mode, it is possible to introduce state and control vectors, respectively $\tx{x}_\tx{drv}$ and $\tx{u}_\tx{drv}$, with respect to $s$, as
\begin{align*}
    \tx{x}_\tx{drv}(s)=\begin{bmatrix} E(s)\\ \tx{soc}^\tx{drv}(s)\\T^\tx{drv}_\tx{b}(s) \end{bmatrix}, \ \tx{u}_\tx{drv}(s)=\begin{bmatrix} P^\tx{b,drv}_\tx{hvch}(s)\vspace{0.15cm}\\ P^\tx{b,drv}_\tx{hvac}(s)\vspace{0.15cm}\\
    a_\tx{t}(s) \end{bmatrix}.
\end{align*}
Accordingly, the relation between the time and space derivatives is given as
\begin{align}
    \frac{\tx{d}\tx{x}_\tx{drv}(t)}{\tx{d}t}=v(s)\frac{\tx{d}\tx{x}_\tx{drv}(s)}{\tx{d}s},\quad t\in\mathcal{T}_\tx{drv}, s\in\mathcal{S}_\tx{drv},\label{eq:dt2ds}
\end{align}
where $\mathcal{S}_\tx{drv}$ is a set including driving distance instances.

Following \eqref{eq:dt2ds}, the longitudinal dynamics \eqref{eq:long_dyn_t} is now described in the space coordinate $s$, as
\begin{align}
  \frac{\tx{d}E(s)}{\tx{d}s}=a_\tx{t}(s)-c_\tx{a}E(s)-a_\alpha(s), \label{eq:long_dyn_s}
\end{align}
where $\frac{\tx{d}E}{\tx{d}s}=\nobreak v\frac{\tx{d}v}{\tx{d}s}$ represents longitudinal acceleration in $s$ domain, and the coefficient $c_\tx{a}=\rho_\tx{a}c_\tx{d}A_\tx{f}/m$ contains the air drag related factors.

Using the relations \eqref{eq:v2E} and \eqref{eq:dt2ds}, the dynamical change of battery SoC with respect to $s$ is given by
\begin{align}
\frac{\tx{d}\tx{soc}^\tx{drv}(s)}{\tx{d}s}=-\frac{P^\tx{drv}_\tx{b}(s)}{C_\tx{b}U_\tx{oc}(\tx{soc}^\tx{drv}(s))\sqrt{2E(s)}}.\label{eq:soc_dyn_s}   
\end{align}

Similarly, the position dependent dynamical change of the battery pack temperature is given by
\begin{align}
\begin{split}
&\frac{\tx{d}T_\tx{b}^\tx{drv}(s)}{\tx{d}s}=\frac{1}{c_\tx{p}m_\tx{b}\sqrt{2E(s)}}\big(Q^\tx{gen}_\tx{pass}(\cdot)+Q_\tx{act}(P_\tx{hvch}^\tx{b,drv}(s),P_\tx{hvac}^\tx{b,drv}(s))\\
&\hspace{3cm}+Q_\tx{exh}(T^\tx{drv}_\tx{b}(s),T_\tx{amb}(s),v(s))\big).
\end{split}\label{eq:tb_dyn_s}
\end{align}

The power balance equation \eqref{eq:pt_Peq} can also be summarized throughout the driving mode, as
\begin{align}
\begin{split}
&P_\tx{b}^\tx{drv}(s)=R(T^\tx{drv}_\tx{b}(s))\frac{\big(P_\tx{b}^\tx{drv}(s)\big)^2}{U_{\tx{oc}}^2(\tx{soc}^\tx{drv}(s))}+P^\tx{drv}_\tx{prop}(v(s),a_\tx{t}(s))\\
&\hspace{1.25cm}+P^\tx{b,drv}_\tx{hvch}(s)+P^\tx{b,drv}_\tx{hvac}(s)+P^{\tx{c}}_{\tx{hvch}}(s)+P^\tx{drv}_{\tx{aux}}(s).\label{eq:peqs}
\end{split}   
\end{align}

%Thus, $E$, $T^\tx{drv}_\tx{b}$ and $\tx{soc}^\tx{drv}$ are the state variables, and $P^\tx{b,drv}_\tx{hvch}$, $P^\tx{b,drv}_\tx{hvac}$, $P^\tx{drv}_\tx{b}$ and $a_\tx{t}$ are the control inputs, that are describing the dynamical system during driving mode. %It is possible to introduce state and control vectors, respectively $\tx{x}_\tx{drv}$ and $\tx{u}_\tx{drv}$, as
% \begin{align*}
%     \tx{x}_\tx{drv}(s)=\begin{bmatrix} E(s)\\ \tx{soc}^\tx{drv}(s)\\T^\tx{drv}_\tx{b}(s) \end{bmatrix}, \ \tx{u}_\tx{drv}(s)=\begin{bmatrix} P^\tx{b,drv}_\tx{hvch}(s)\vspace{0.15cm}\\ P^\tx{b,drv}_\tx{hvac}(s)\vspace{0.15cm}\\
%     P^\tx{drv}_\tx{b}(s)\\a_\tx{t}(s) \end{bmatrix},
% \end{align*}
%that include the state variables and control inputs. 
The governing dynamics during driving mode can be summarized as
\begin{align*}
    \frac{\tx{d}\tx{x}_\tx{drv}(s)}{\tx{d}s}=f_\tx{drv}(\tx{x}_\tx{drv}(s),\tx{u}_\tx{drv}(s),s),
\end{align*}
where $f_\tx{drv}$ is a vector function including nonlinear scalar functions illustrating each state variable's dynamical change, according to \eqref{eq:long_dyn_s}-\eqref{eq:tb_dyn_s}. We also define a vector $\tx{x}^\tx{ts}_\tx{drv}$, as
\begin{align*}
    \tx{x}^\tx{ts}_\tx{drv}(s)=\begin{bmatrix} \tx{soc}^\tx{drv}(s)\\T^\tx{drv}_\tx{b}(s) \end{bmatrix},
\end{align*}
which will be used later for describing the transition between the modes.

The performance function during driving mode includes the penalty on trip time, as
\begin{align}
    J_\tx{drv}(\cdot)=\int_{s\in \mathcal{S}_\tx{drv}}^{} \frac{c_\tx{t,trip}}{\sqrt{2E(s)}}\tx{d}s,
    \label{eq:objdrv}
\end{align}
which is directly obtained from the trip time to travel distance transformation, i.e. $\int c_\tx{t,trip}\tx{d}t=\int c_\tx{t,trip}/\sqrt{2E(s)} \,\tx{d}s.$ The set $\mathcal{S}_\tx{drv}$ includes the driving distance instances.

\subsection{Charging Mode: Dynamics and Performance Function}\label{subsec:chg_mode}
Governing dynamics during charging mode corresponds to the dynamical changes of battery temperature and SoC. The state variables and control inputs of charging mode for $i\in\mathcal{I}$ are stacked, respectively, in vectors $\tx{x}^{i}_\tx{chg}$ and $\tx{u}^{i}_\tx{chg}$, as 
\begin{align*}
    \tx{x}^{i}_\tx{chg}(\tau^i)=\begin{bmatrix} \tx{soc}^{i\tx{,chg}}(\tau^i)\\T^{i\tx{,chg}}_\tx{b}(\tau^i) \end{bmatrix}, \ \tx{u}^{i}_\tx{chg}(\tau^i)=\begin{bmatrix} P^{i\tx{,b,chg}}_\tx{hvch}(\tau^i)\vspace{0.15cm}\\ P^{i\tx{,b,chg}}_\tx{hvac}(\tau^i)\vspace{0.15cm}\\
    P^{i}_\tx{grid}(\tau^i) \end{bmatrix}, \ i\in \mathcal{I}.
\end{align*}
Also, the charging time associated with each charging station is considered as a scalar variable, which is optimized simultaneously with the optimal state and control trajectories of both driving and charging modes. % and included in a vector $\tx{T}_\tx{chg}$, as
% \begin{align*}
%     \tx{T}_\tx{chg}=\begin{bmatrix}t^{1}_\tx{chg}\vspace{0.15cm}\\t^{2}_\tx{chg}\vspace{0.15cm}\\\vdots\vspace{0.15cm}\\t^{N_\tx{chg}}_\tx{chg}\vspace{0.15cm}\\\end{bmatrix},
% \end{align*}
According to \eqref{eq:t2tau}, the relation between the time derivative and the derivative with respect to $\tau^{i}\in[0,1]$, $i\in\mathcal{I}$, is
\begin{align}
    \frac{\tx{d}\tx{x}^{i}_\tx{chg}(t)}{\tx{d}t}=\frac{1}{t^{i}_\tx{chg}}\frac{\tx{d}\tx{x}^{i}_\tx{chg}(\tau^{i})}{\tx{d}\tau^{i}},\quad t\in\mathcal{T}^{i}_\tx{chg}, s(t)=s^{i}_\tx{chg}.\label{eq:dt2dtau}
\end{align}

Following \eqref{eq:dt2dtau}, the dynamical variation of battery SoC with respect to $\tau^i$ for $i\in\mathcal{I}$ is given by
\begin{align}
\frac{\tx{d}\tx{soc}^{i\tx{,chg}}(\tau^i)}{\tx{d}\tau^i}=-\frac{t^{i}_\tx{chg}P^{i\tx{,chg}}_\tx{b}(\tau^i)}{C_\tx{b}U_\tx{oc}(\tx{soc}^{i\tx{,chg}}(\tau^i))}.\label{eq:soc_dyn_tau}   
\end{align}

Similarly, the $\tau^i$ dependent dynamical change of the battery pack temperature for $i\in\mathcal{I}$ is given by
\begin{align}
\begin{split}
&\frac{\tx{d}T_\tx{b}^{i\tx{,chg}}(\tau^i)}{\tx{d}\tau^i}=\frac{t^{i}_\tx{chg}}{c_\tx{p}m_\tx{b}}\big(Q^\tx{gen}_\tx{pass}(\cdot)+Q_\tx{act}(P_\tx{hvch}^{i\tx{,b,chg}}(\tau^i),P_\tx{hvac}^{i\tx{,b,chg}}(\tau^i))\\
&\hspace{3cm}+Q_\tx{exh}(T^{i\tx{,chg}}_\tx{b}(\tau^i),T_\tx{amb}(\tau^i))\big).
\end{split}\label{eq:tb_dyn_tau}
\end{align}

For $i\in\mathcal{I}$, the power balance equation \eqref{eq:pt_Peq} during the charging modes is
\begin{align}
\begin{split}
P^{i}_\tx{grid}(\tau^i)&+P^{i\tx{,chg}}_\tx{b}(\tau^i)=R(T^{i\tx{,chg}}_\tx{b}(\tau^i))\frac{\big(P_\tx{b}^{i\tx{,chg}}(\tau^i)\big)^2}{U_{\tx{oc}}^2(\tx{soc}^{i\tx{,chg}}(\tau^i))}\\
&+P^{i\tx{,b,chg}}_\tx{hvch}(\tau^i)+P^{i\tx{,b,chg}}_\tx{hvac}(\tau^i)+P^{i\tx{,chg}}_{\tx{aux}}(\tau^i). 
\end{split}\label{eq:peqtau}    
\end{align}
Note that propulsion power is equal to zero during charging in \eqref{eq:peqtau}. Also, the power demand for heating the cabin compartment during charging is assumed to be zero in \eqref{eq:peqtau}, %since charging time can be long enough that
which is reasonable for the case when the driver/passengers stay outside the vehicle during charging. 

% The state variables and control inputs of charging mode for $i\in\mathcal{I}$ are stacked, respectively, in vectors $\tx{x}^{i}_\tx{chg}$ and $\tx{u}^{i}_\tx{chg}$, as 
% \begin{align*}
%     \tx{x}^{i}_\tx{chg}(\tau^i)=\begin{bmatrix} \tx{soc}^{i\tx{,chg}}(\tau^i)\\T^{i\tx{,chg}}_\tx{b}(\tau^i) \end{bmatrix}, \ \tx{u}^{i}_\tx{chg}(\tau^i)=\begin{bmatrix} P^{i\tx{,b,chg}}_\tx{hvch}(\tau^i)\vspace{0.15cm}\\ P^{i\tx{,b,chg}}_\tx{hvac}(\tau^i)\vspace{0.15cm}\\
%     P^{i,\tx{chg}}_\tx{b}(\tau^i)\vspace{0.15cm}\\
%     P^{i}_\tx{grid}(\tau^i) \end{bmatrix}, \ i\in \mathcal{I}.
% \end{align*}
The governing dynamics during charging mode for $i\in\mathcal{I}$ can be summarized as
\begin{align*}
    \frac{\tx{d}\tx{x}^{i}_\tx{chg}(\tau^i)}{\tx{d}\tau^i}=f_\tx{chg}(\tx{x}^{i}_\tx{chg}(\tau^i),\tx{u}^{i}_\tx{chg}(\tau^i),t^{i}_\tx{chg},\tau^i),
\end{align*}
where $f_\tx{chg}$ is a vector function including nonlinear scalar functions describing each state variable's dynamical variation, according to \eqref{eq:soc_dyn_tau} and \eqref{eq:tb_dyn_tau}.

The performance function associated with charging mode for $i\in\mathcal{I}$, is the compromise among charging energy cost, charging time and charger occupying time cost, as
\begin{align}
\begin{split}
&J_\tx{chg}(\cdot)=\sum_{i=1}^{N_\tx{chg}}\Big(t_\tx{chg}^{i}\int_{0}^{1}
\big(c_\tx{t,trip}+c^{i}_\tx{e}P^{i}_\tx{grid}(\tau^i)\big)\tx{d}\tau^i\\
&\hspace{3.5cm}+c^{i}_\tx{T}\max\big(0,t^{i}_\tx{chg}-T^{i}\big)\Big).
\end{split}\label{eq:objchg}
\end{align}

\subsection{Hybrid Dynamical System Formulation}\label{subsec:pf_hybrid}
The hybrid dynamical system's formulation for $i\in\mathcal{I}$, can now be summarized as
{\allowdisplaybreaks
\begin{subequations} \label{eq:phyb}
\begin{align}
&\min_{\tx{u}_\tx{drv}(s),\tx{u}^{i}_\tx{chg}(\tau^i),t^i_\tx{chg}}J_\tx{drv}(\cdot)+J_\tx{chg}(\cdot)\\
&\text{for $\tau^i\in[0,1]$ subject to:} \nonumber\\
&\frac{\tx{d}\tx{x}_\tx{drv}(s)}{\tx{d}s}=f_\tx{drv}(\tx{x}_\tx{drv}(s),\tx{u}_\tx{drv}(s),s), \quad s\in\mathcal{S}_\tx{drv}\label{eq:phyb_sdyn}\\
&\frac{\tx{d}\tx{x}^{i}_\tx{chg}(\tau^i)}{\tx{d}\tau^i}=f_\tx{chg}(\tx{x}^{i}_\tx{chg}(\tau^i),\tx{u}^{i}_\tx{chg}(\tau^i),\tau^i), \quad s\in s^{i}_\tx{chg}\label{eq:phyb_taudyn}\\
&g_\tx{drv}(\tx{x}_\tx{drv}(s),\tx{u}_\tx{drv}(s),s)\leq 0, \quad s\in\mathcal{S}_\tx{drv}\label{eq:phyb_gscns}\\
&g_\tx{chg}(\tx{x}^{i}_\tx{chg}(\tau^i),\tx{u}^{i}_\tx{chg}(\tau^i),\tau^i)\leq 0, \quad s\in s^{i}_\tx{chg}\label{eq:phyb_gtaucns}\\
&\tx{x}_\tx{drv}(s)\in\mathcal{X}_\tx{drv}(s),\quad \tx{u}_\tx{drv}(s)\in\mathcal{U}_\tx{drv}(s),\quad s\in\mathcal{S}_\tx{drv}\label{eq:phyb_drv_bcns}\\
&\tx{x}^{i}_\tx{chg}(\tau^i)\in\mathcal{X}^{i}_\tx{chg}(\tau^i),\quad \tx{u}^{i}_\tx{chg}(\tau^i)\in\mathcal{U}^{i}_\tx{chg}(\tau^i),\quad s\in s^{i}_\tx{chg}\label{eq:phyb_chg_bcns}\\
&\tx{x}^i_\tx{chg}(0)=\tx{x}^\tx{ts}_\tx{drv}(s^{i}_\tx{chg})\label{eq:drv2chg}\\
&\tx{x}^\tx{ts}_\tx{drv}(s_\tx{chg}^{i^{+}})=\tx{x}^{i}_\tx{chg}(1)\label{eq:chg2drv}\\
&\tx{x}_\tx{drv}(s_0)\in\mathcal{X}_\tx{drv0}, \quad \tx{x}_\tx{drv}(s_\tx{f})\in\mathcal{X}_\tx{drvf}
\end{align}
\end{subequations}}%
where $t_\tx{chg}^i$ is treated as a design parameter, $s_\tx{chg}^{i^{+}}$ is an instance where the vehicle is leaving the charging station,  $g_\tx{drv}$ and $g_\tx{chg}$ denote the system general constraints, respectively during driving and charging modes, including the bounds on battery power and traction acceleration, as
{\allowdisplaybreaks
\begin{subequations}
\begin{align}
&g_\tx{drv}(\cdot)=\left\{
    \begin{array}{c}
         \mbox{$P^{\min}_{\tx{b,chg}}(\tx{soc}^\tx{drv}(s),T^\tx{drv}_\tx{b}(s))-P^\tx{drv}_\tx{b}(s)$},\vspace{0.15cm}\\    
         \mbox{$P^\tx{drv}_\tx{b}(s)-P^{\max}_{\tx{b,dchg}}(\tx{soc}^\tx{drv}(s),T^\tx{drv}_\tx{b}(s))$},\vspace{0.15cm}\\
         \mbox{$a_{\min}(E(s))-a_\tx{t}(s)$},\vspace{0.15cm}\\         
         \mbox{$a_\tx{t}(s)-a_{\max}(E(s))$}
    \end{array}
\right\}\label{eq:gdrv}\\
&g_\tx{chg}(\cdot)=\left\{
    \begin{array}{c}
         \mbox{$P^{\min}_{\tx{b,chg}}(\tx{soc}^{i,\tx{chg}}(\tau^i)),T^{i,\tx{chg}}_\tx{b}(\tau^i))-P^{i,\tx{chg}}_\tx{b}(\tau^i)$}.
    \end{array}
\right\}\label{eq:gchg}
\end{align}
\end{subequations}}%
Also, $\mathcal{X}_\tx{drv}$ and $\mathcal{X}_\tx{chg}$ denote the feasible sets of state variables, and $\mathcal{U}_\tx{drv}$ and $\mathcal{U}_\tx{chg}$ represent the feasible sets of control inputs for each mode. %, as
% {\allowdisplaybreaks
% \begin{subequations}
% \begin{align}
% &\mathcal{X}_\tx{drv}(s)=\left\{
%     \begin{array}{c}
%          \mbox{$[T_\tx{b}^{\min}(s),T_\tx{b}^{\max}(s)]$},\vspace{0.15cm}\\
%          \mbox{$[\tx{soc}_{\min}(s),\tx{soc}_{\max}(s)]$},\vspace{0.15cm}\\
%          \mbox{$\frac{1}{2}[v_{\min}(s),v_{\max}(s)]$}
%     \end{array}
% \right\}\label{eq:bxdrv}\\
% &\mathcal{U}_\tx{drv}(s)=\left\{
%     \begin{array}{c}
%          \mbox{$[0,P^{\max}_{\tx{hvch}}-P^{\tx{c}}_{\tx{hvch}}(s)]$},\vspace{0.15cm}\\
%          \mbox{$[0,P^{\max}_{\tx{hvac}}]$}
%     \end{array}
% \right\}\label{eq:budrv}
% \end{align}
% \end{subequations}}%
% {\allowdisplaybreaks
% \begin{subequations}
% \begin{align}
% &\mathcal{X}_\tx{chg}(\tau^i)=\left\{
%     \begin{array}{c}
%          \mbox{$[T_\tx{b}^{\min}(\tau^i),T_\tx{b}^{\max}(\tau^i)]$},\vspace{0.15cm}\\
%          \mbox{$[\tx{soc}_{\min}(\tau^i),\tx{soc}_{\max}(\tau^i)]$}
%     \end{array}
% \right\}\label{eq:bxchg}\\
% &\mathcal{U}_\tx{chg}(\tau^i)=\left\{
%     \begin{array}{c}
%          \mbox{$[0,P^{\max}_{\tx{hvch}}]$},\vspace{0.15cm}\\
%          \mbox{$[0,P^{\max}_{\tx{hvac}}]$},\vspace{0.15cm}\\
%          \mbox{$[0,t_\tx{chg}^{\max}]$}
%     \end{array}
% \right\}\label{eq:buchg}
% \end{align}
% \end{subequations}}%
Furthermore, $\mathcal{X}_\tx{drv0}$ and $\mathcal{X}_\tx{drvf}$ denote allowed initial states at $\tx{s}_\tx{0}$, and target states at $\tx{s}_\tx{f}$, respectively. The constraints \eqref{eq:drv2chg} and \eqref{eq:chg2drv} denote the transition between the modes. Thus, the battery temperature and SoC at the arrival of charging station must be equal to the corresponding variables when charging begins. Similarly, the battery temperature and SoC when charging is just finished must be equal to the corresponding variables when the vehicle resumes its drive.

\section{Results}\label{sec:res}
In this section, simulation results are provided for the scenario illustrated in Fig.~\ref{fig:scenario}. Within the simulations, we evaluate the performance of the proposed algorithm to achieve optimal thermal management, eco-driving, and charging of a BEV in the presence of existing constraints. The simulation setup is given in Section~\ref{subsec:setup}.

\subsection{Simulation Setup}\label{subsec:setup}
The simulations are performed for a BEV over a \SI{440}{km} long road with a hilly terrain. The BEV starts its mission with \SI{80}{\%} SoC and cold battery, where ambient temperature is also low during the vehicle's entire mission, i.e. $T_{\tx{b0}}(s_0)=T_\tx{amb}(s)=\SI{-10}{^\circ C}, s\in[s_0,s_\tx{f}]$. Followed by the constant ambient temperature, the HVCH power demand for heating the cabin compartment during the vehicle's driving mode is also a fixed value. As the driving distance is greater than the vehicle's electric range, one intermediate charging station is visited at $s=\SI{240}{km}$, and a terminal charging station is also considered at the end of the route. The terminal battery SoC is set to be the same percentage as the initial SoC, i.e. \SI{80}{\%}. Also, the rated grid power provided by the chargers as well as the rated battery charging power are \SI{150}{kW}. Note that the time based cost for occupying the charging spot is not considered in the studied scenario, i.e. $c_\tx{T}=0$. The vehicle and simulation parameters are provided in Table \ref{tab:par}. 

The NLP \eqref{eq:phyb} is discretized using the Runge-Kutta $4^{th}$ order method~\cite{butcher76}, with a distance sampling interval of \SI{2}{km}. Subsequently, the discretized problem is solved in Matlab with the solver IPOPT, using the open source nonlinear optimisation tool CasADi~\cite{andersson19}. The optimisation was run on a laptop PC with 6600K CPU at $2.81$GHz and $16$GB RAM, where the solving time is less than a minute.

%According to the simulation results, electrical energy consumption cost during the driving mode is \SI{180}{SEK}, and total charging electricity cost is \SI{565}{SEK}.

\begin{table}
\begin{center}
\caption{Vehicle and Simulation Parameters} 
\label{tab:par}
\begin{tabular}{l l}
\hline
Gravitational acceleration & $g=\SI{9.81}{m/s^2}$\vspace{0.05cm}\\
Air density & $\rho_\tx{a}=\SI{1.29}{kg/m^3}$\vspace{0.05cm}\\
Vehicle frontal area & $A_\tx{f}=\SI{1.36}{m^2}$\vspace{0.05cm}\\
Rolling resistance coefficient & $c_\tx{r}=\SI{0.013}{}$\vspace{0.05cm}\\
Total vehicle mass & $m=\SI{2200}{kg}$\vspace{0.05cm} \\
Aerodynamic drag coefficient & $c_\tx{d}=\SI{0.6}{}$\vspace{0.05cm}\\
Maximum battery capacity & $C_\tx{p}=\SI{200}{Ah}$\vspace{0.05cm}\\ 
Product of specific heat capacity and total vehicle mass & $c_\tx{p}m_\tx{b}=\SI{375}{kJ/(K)}$\vspace{0.05cm}\\ 
%Total battery mass & $m_\tx{b}=\SI{372}{kg}$\vspace{0.05cm}\\ 
Route length & $\SI{480}{km}$\vspace{0.05cm} 
\\
Distance sampling interval & $\SI{2}{km}$\vspace{0.05cm} \\
Number of charging along the route  & $N_\tx{chg}=\SI{2}{}$\vspace{0.05cm} \\
%Electrical energy cost while driving & %$c_\tx{e,drv}=\SI{1.65}{SEK/kWh}$\vspace{0.05cm} \\
Electrical energy cost while charging & $c_\tx{e}=\SI{5}{SEK/kWh}$\vspace{0.05cm} \\
Charger rated power & $P^{\max}_\tx{grid}=\SI{150}{kW}$\vspace{0.05cm} \\
Auxiliary load & $P_\tx{aux}=\SI{0.5}{kW}$\vspace{0.05cm} \\
%Max. deliverable HVCH power & $P^{\max}_{\tx{hvch}}=\SI{7}{kW}$\vspace{0.05cm} \\
%Max. deliverable HVAC power & $P^{\max}_{\tx{hvac}}=\SI{7}{kW}$\vspace{0.05cm} \\
HVCH power for heating cabin & $P^\tx{c}_{\tx{hvch}}=\SI{1.5}{kW}$ \vspace{0.05cm}\\
HVCH power to heat rate efficiency & $\eta_{\tx{hvch}}=\SI{87}{\%}$\vspace{0.05cm} \\
HVAC power to heat rate efficiency & $\eta_{\tx{hvac}}=\SI{87}{\%}$\vspace{0.05cm} \\
Initial battery temperature & $T_{\tx{b0}}=\SI{-10}{^\circ C}$\vspace{0.05cm} \\
Ambient temperature & $T_\tx{amb}=\SI{-10}{^\circ C}$\vspace{0.05cm} \\
Initial battery state of charge & $\tx{soc}_0=\SI{80}{\%}$\vspace{0.05cm} \\
Terminal battery state of charge & $\tx{soc}_\tx{f}=\SI{80}{\%}$\vspace{0.05cm} \\
Minimum speed limit &
$v_\tx{min}=\SI{65}{km/h}$\vspace{0.05cm} \\
Maximum speed limit &
$v_\tx{max}=\SI{110}{km/h}$\vspace{0.05cm} \\
\hline
\end{tabular}
\end{center}
\end{table}

\begin{figure}[t!]
 \centering
 \includegraphics[width=0.9\linewidth]{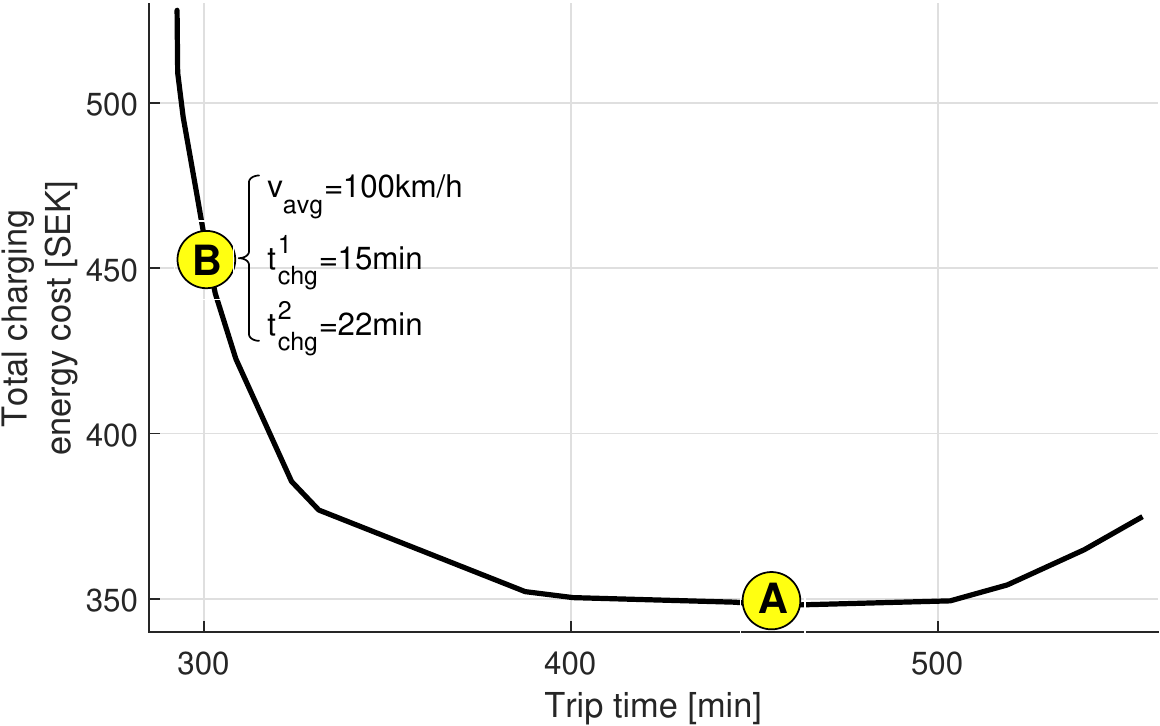}
  \caption{\footnotesize Pareto frontier describing the trade-off between total charging energy
  cost versus trip time.}
  \label{fig:pf}
\end{figure}

\begin{figure}[t!]
\centering
\subfigure[Road topography together with vehicle speed profile and speed limits.]{
 \includegraphics[width=.92\linewidth]{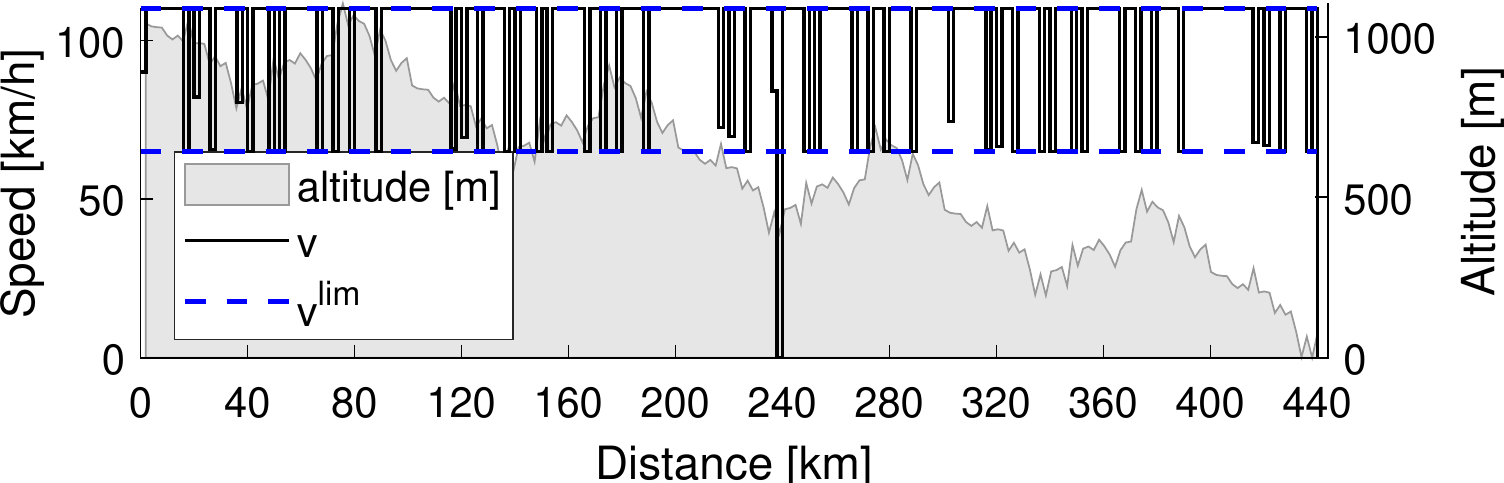}
\label{fig:v}
}
\subfigure[Battery state of charge trajectory together with its bounds.]{

 \includegraphics[width=.9\linewidth]{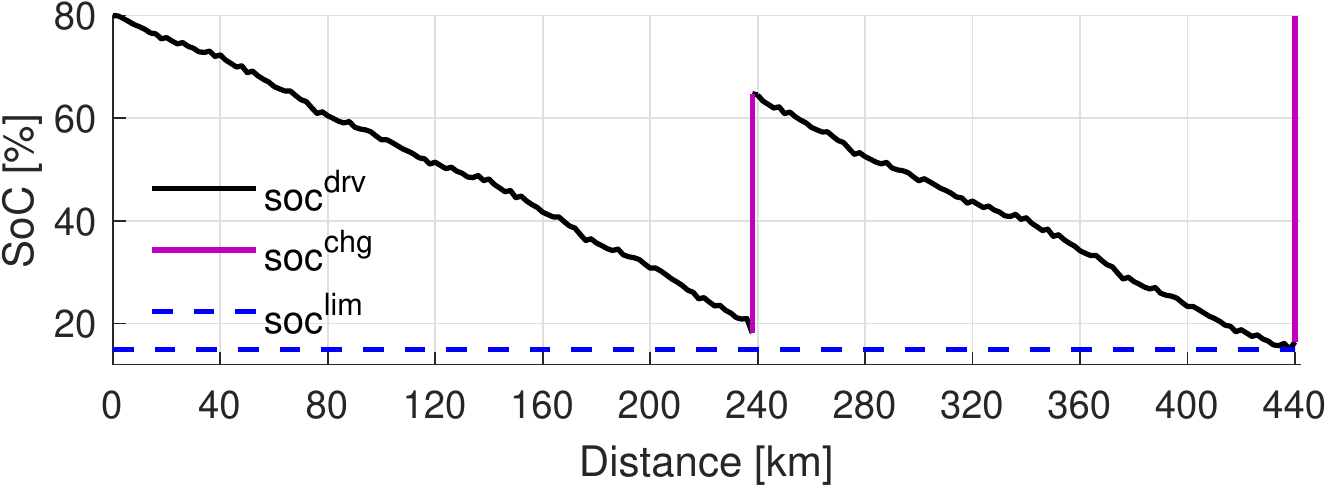}
\label{fig:socdrv}
}
\subfigure[Battery temperature trajectory together with its upper bound and ambient temperature.]{

 \includegraphics[width=.9\linewidth]{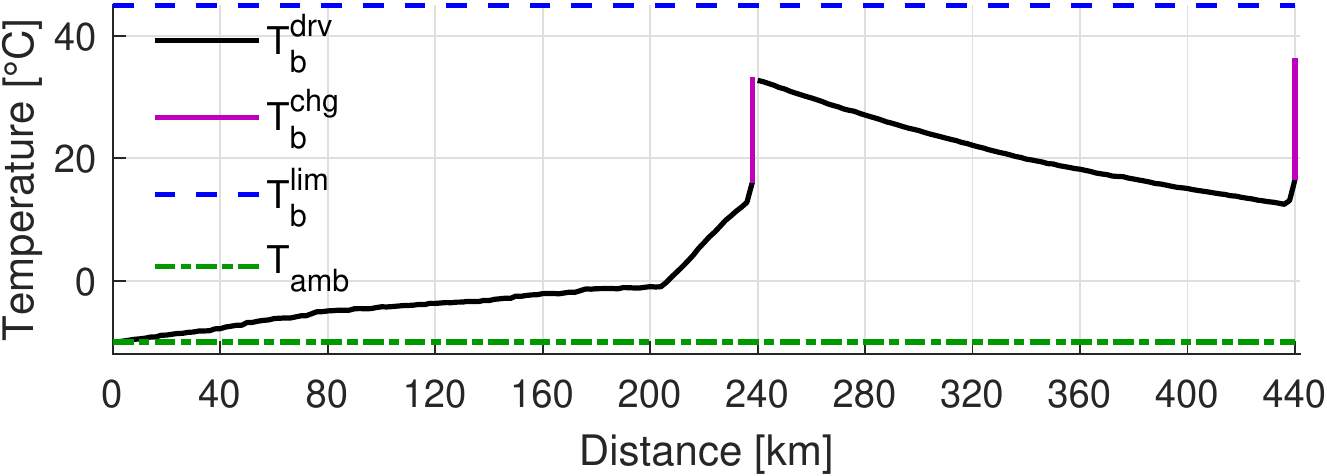}
\label{fig:tbdrv}
}
\subfigure[Trajectories of HVCH and HVAC power for battery heating.]{

 \includegraphics[width=.9\linewidth]{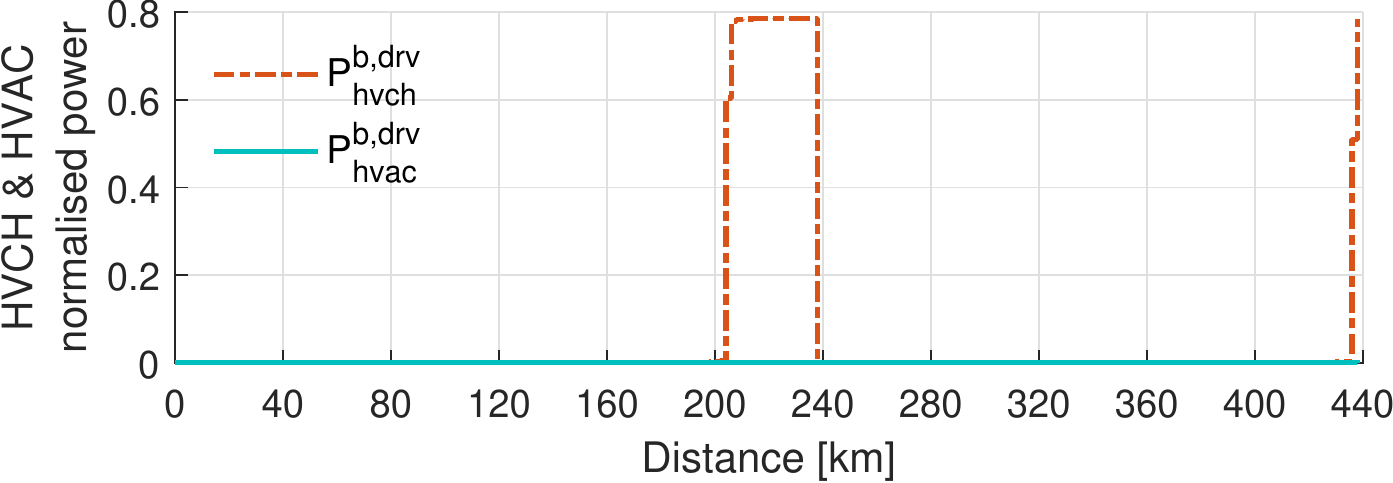}
\label{fig:hvchhvacdrv}
}
\subfigure[Trajectories of battery power and propulsion power together with battery discharge power limit.]{

 \includegraphics[width=.9\linewidth]{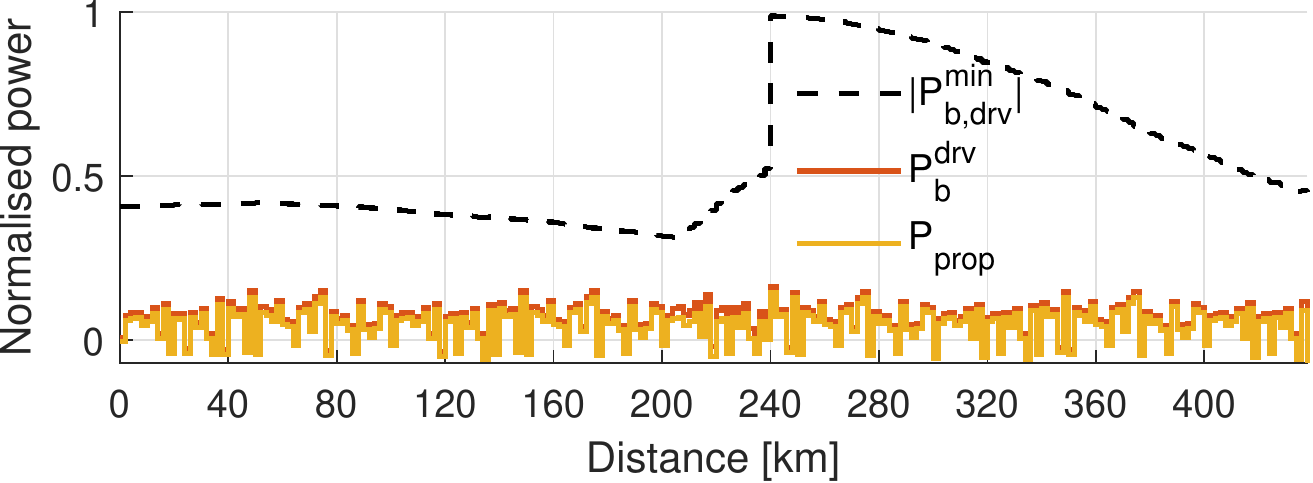}
\label{fig:pbpropdrv}
}
\caption{Case 1; optimal trajectories Vs. travelled distance. The step changes in battery temperature and SoC at $s=\SI{240}{km}$ and $s=\SI{440}{km}$ in (a) and (b), denote the increase in the corresponding variables during charging mode.}
\label{fig:drv}
\end{figure}

\subsection{Energy Efficiency Vs. Time}\label{subsec:evst}
To investigate the trade-off between total charging energy cost versus trip time, the Pareto frontier is derived, as shown in Fig.~\ref{fig:pf}, where the total charging cost includes the electrical energy cost during the intermediate and terminal charging modes. Also, the trip time covers the driving and charging times. The driving time variations can be characterised as changing the vehicle's average speed. The demonstrated Pareto frontier provides a wide range of choices for various types of car users to customise their trip. In Fig.~\ref{fig:pf}, point A denotes the vehicle's most energy efficient trip, where $c_\tx{t,trip}=0$. The trip time can be increased further by letting $c_\tx{t,trip}$ be negative, where this leads to an increase in the energy cost. Thus, there is a low average speed $v_\tx{avg}$ threshold, here about $v_\tx{avg}\approx\SI{70}{km/h}$, below which the increased time of accumulating powertrain losses prevails the benefit of reduced air drag. Point B in Fig.~\ref{fig:pf} corresponds to a more time efficient trip, i.e. $v_\tx{avg}\approx\SI{100}{km/h}$, compared to the case in point A. 

In the rest of the paper we will only consider the vehicle's operation in point B. In this point, Case 1, i.e. with active heating/cooling, is compared to Case 2, i.e. without active heating/cooling, to evaluate the impact of \textit{battery preconditioning} on the charging time and energy cost. Battery preconditioning is characterized as bringing the battery temperature to (or closer to) its desired range, where discharging/charging power availability is increased considerably.

\begin{table}
\caption{Charging Cost Vs. Trip Time}\vspace{-0.25cm}
\label{tab:costvst}
\begin{center}
\begin{tabular}{ c|c|c } 
%\hline
Case & Trip Time (Total Chg. Time) [min] & Chg. Cost [SEK]\\
\hline
\multirow{1}{4.75em}{Case 1} & 294 (37) & 453 \\ 
\multirow{1}{4.75em}{Case 2} & 323 (66) & 444 \\
%\hline
\end{tabular}
\end{center}
\end{table}

\subsection{Case 1: Time Efficient Trip with Active Heating/Cooling}\label{subsec:acthc}
Here, the results are categorized into the optimal trajectories versus travelled distance, and versus charging time during the intermediate and terminal charging events. Total charging cost and trip time are given in Table \ref{tab:costvst}.

\subsubsection{Optimal Trajectories Vs. Travelled Distance}\label{subsubsection:drv}
Optimal vehicle speed profile together with the speed limits and road topography are depicted in Fig.~\ref{fig:v}, where the zero speed values at travel distances $s=\SI{240}{km}$ and $s=\SI{440}{km}$ resemble the vehicle stops at the charging stations. The battery depletes gradually as the vehicle continues its drive, where at the arrival of the charging stations at $s=\SI{240}{km}$ and $s=\SI{440}{km}$, the SoC levels are about $\SI{20}{\%}$ and $\SI{15}{\%}$, respectively as demonstrated in Fig.~\ref{fig:socdrv}. The battery temperature increases primarily due to only the passive heat generation resources, i.e. Joule heat and ED losses, from $s=\SI{0}{km}$ to $s=\SI{205}{km}$, according to Fig.~\ref{fig:tbdrv}. Later, the HVCH, jointly with the passive heat resources, further raise the battery temperature (from $s=\SI{205}{km}$ to $s=\SI{240}{km}$, and from $s=\SI{435}{km}$ to $s=\SI{440}{km}$). Such battery temperature increase by the HVCH demonstrates the battery preconditioning. As shown in Fig.~\ref{fig:pbchglim}, the charging battery power availability is high for low SoC and high battery temperature region. This leads to a reduced charging time, but higher charging cost instead. Note that the decreasing battery temperature from $s=\SI{240}{km}$ to $s=\SI{435}{km}$ is due to an increased heat transfer to the ambient air, as the temperature difference between the battery pack and ambient air is large for this distance segment. In the intermediate and terminal charging stations, the battery is charged to $\SI{63}{\%}$ and $\SI{80}{\%}$ SoC levels, respectively. The propulsion power, battery discharge power and its limit are shown in Fig.~\ref{fig:pbpropdrv}. The battery discharge power limit has a step increase at $s=\SI{240}{km}$, due to the steep rise in SoC and battery temperature due to charging.

\subsubsection{Intermediate Charging}\label{subsubsection:intchg}
\begin{figure}[t!]
\centering
\subfigure[Battery state of charge trajectory together with its bounds.]{

 \includegraphics[width=.9\linewidth]{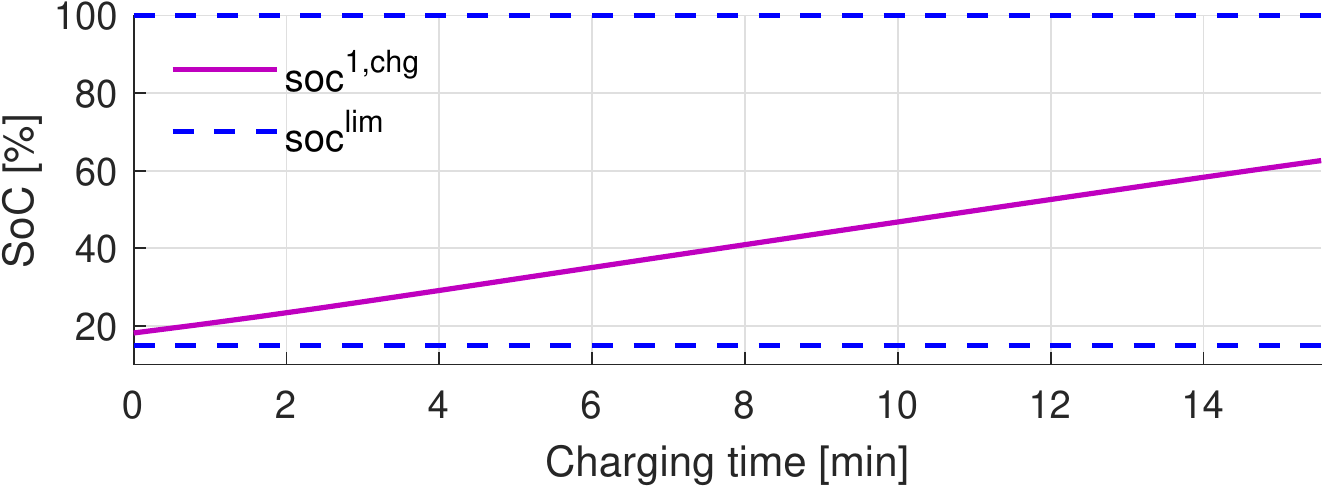}
\label{fig:socchg1}
}
\subfigure[Battery temperature trajectory together with its bounds.]{

 \includegraphics[width=.9\linewidth]{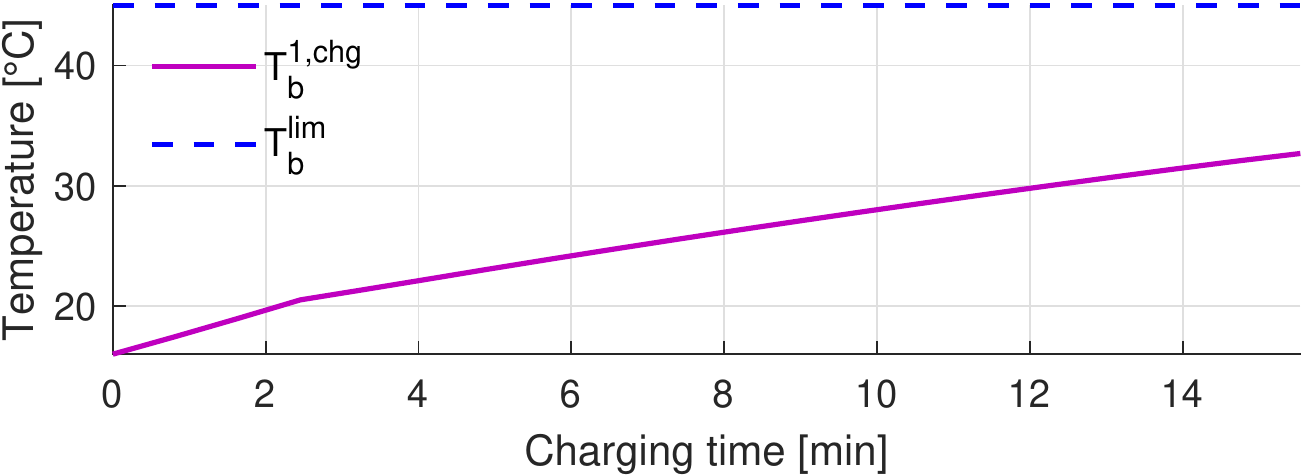}
\label{fig:tbchg1}
}
\subfigure[Trajectories of HVCH and HVAC power for battery heating.]{

 \includegraphics[width=.9\linewidth]{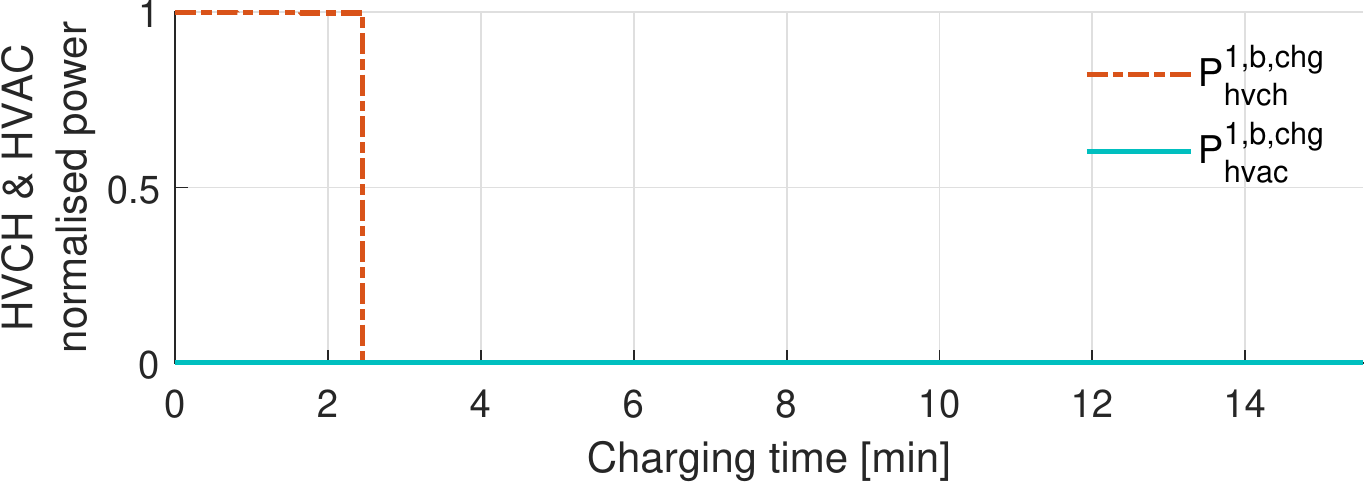}
\label{fig:hvchhvacchg1}
}
\subfigure[Grid power together with absolute values of battery charging power and its limit, versus battery temperature and SoC.]{

 \includegraphics[width=.9\linewidth]{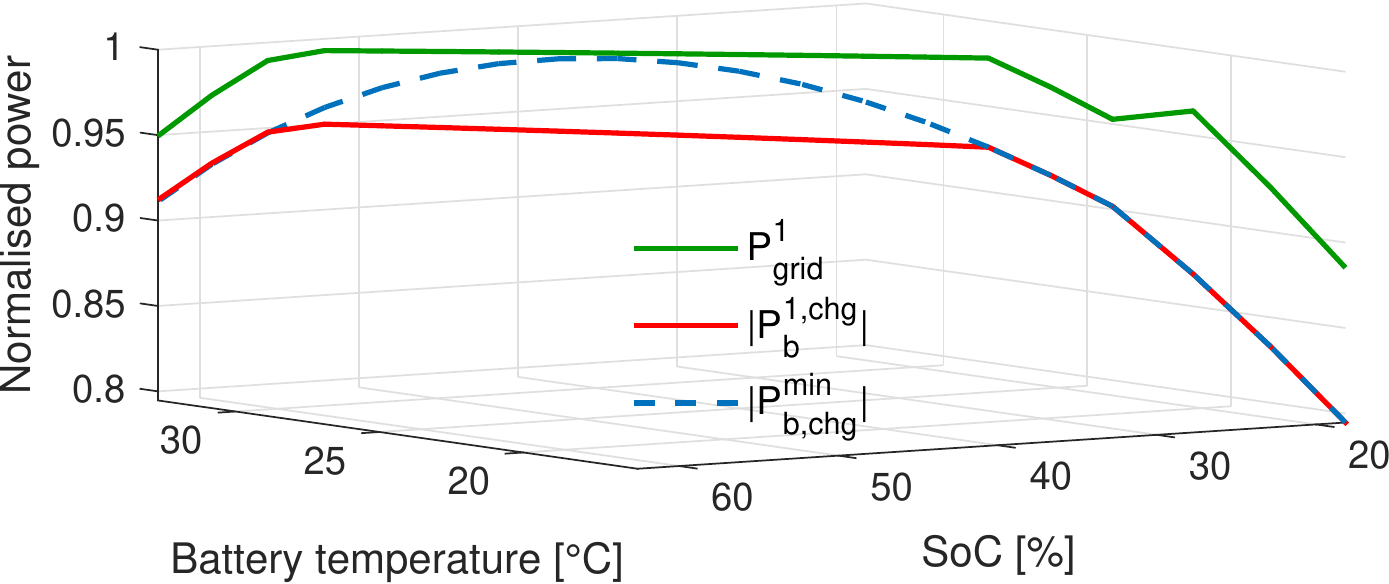}
\label{fig:pbgridtbchg1}
}
\caption{Case 1; optimal trajectories associated with the vehicle's intermediate charging mode at $s=\SI{240}{km}$.}
\label{fig:chg1}
\end{figure}

During the intermediate charging, in addition to the SoC level increase, the battery temperature also rises steadily, as shown in Fig.~\ref{fig:socchg1} and Fig.~\ref{fig:tbchg1}. SoC level throughout the intermediate charging is always in a range with high charging power availability. Also, HVCH stays on for about \SI{2.5}{min} from the beginning of charging, in order to further raise the battery temperature above $\SI{20}{^\circ C}$. This allows charging with high power and for a short time period, which is about $\SI{15}{min}$ here. Fig.~\ref{fig:pbgridtbchg1} illustrates a 3D plot including grid power as well as the absolute values of battery charging power and its limit versus SoC and battery temperature values. The difference between the grid power and battery power is due to the Joule heat losses and the HVCH power demand for heating the battery pack.

\subsubsection{Terminal Charging}\label{subsubsection:trmchg}
\begin{figure}[t!]
\centering
\subfigure[Battery state of charge trajectory together with its bounds.]{

 \includegraphics[width=.9\linewidth]{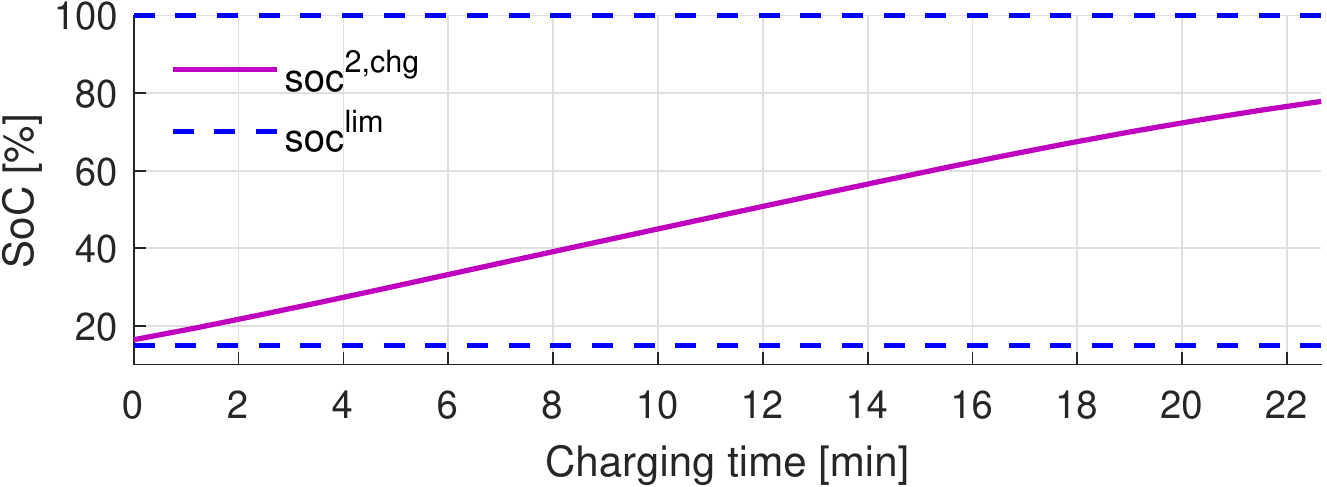}
\label{fig:socchg2}
}
\subfigure[Battery temperature trajectory together with its bounds.]{

 \includegraphics[width=.9\linewidth]{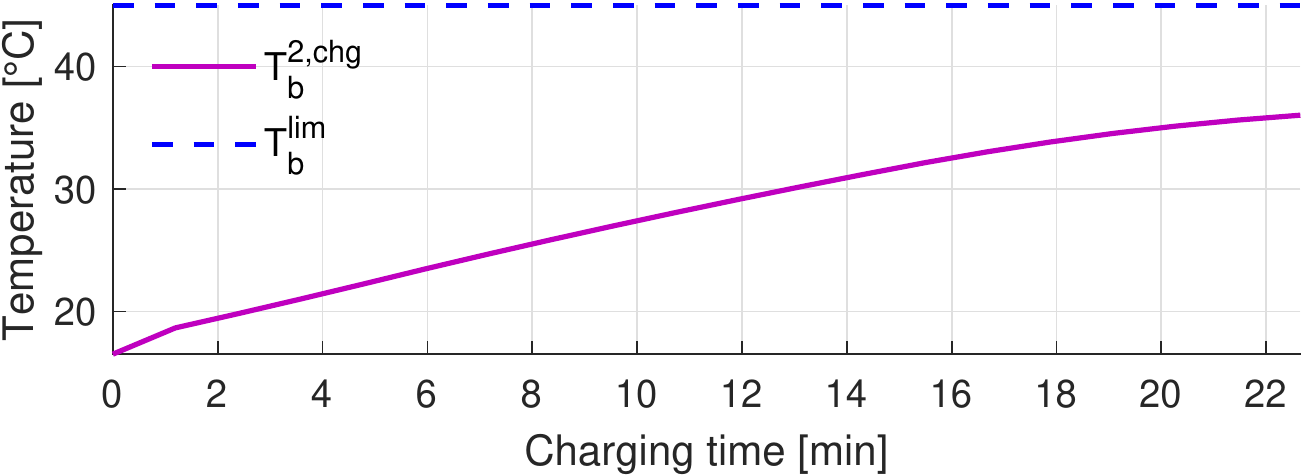}
\label{fig:tbchg2}
}
\subfigure[Trajectories of HVCH and HVAC power for battery heating.]{

 \includegraphics[width=.9\linewidth]{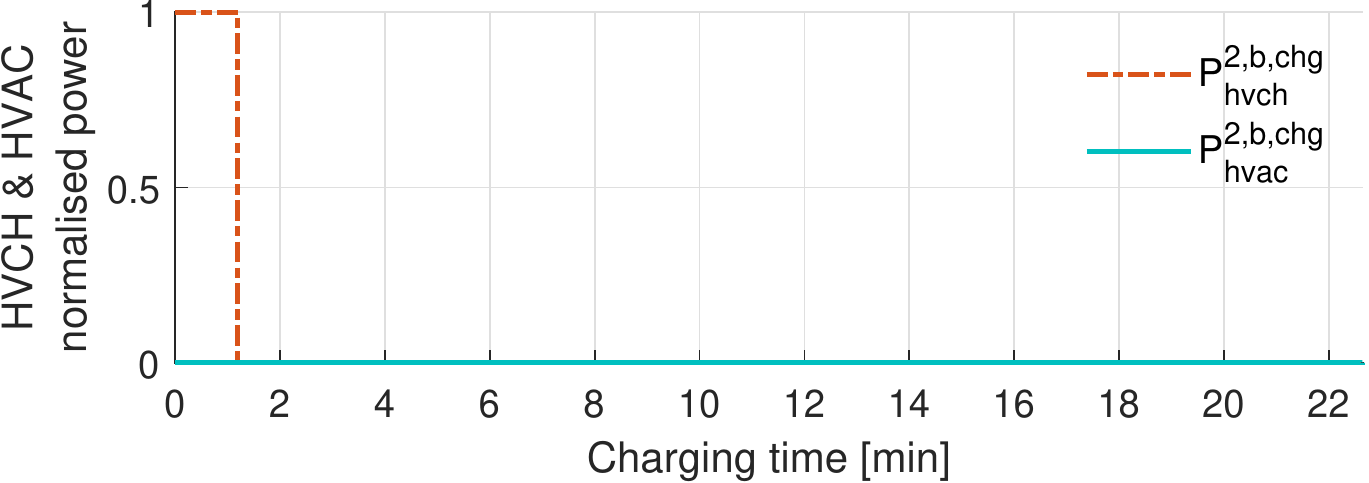}
\label{fig:hvchhvacchg2}
}
\subfigure[Grid power together with absolute values of battery charging power and its limit, versus battery temperature and SoC.]{

 \includegraphics[width=.9\linewidth]{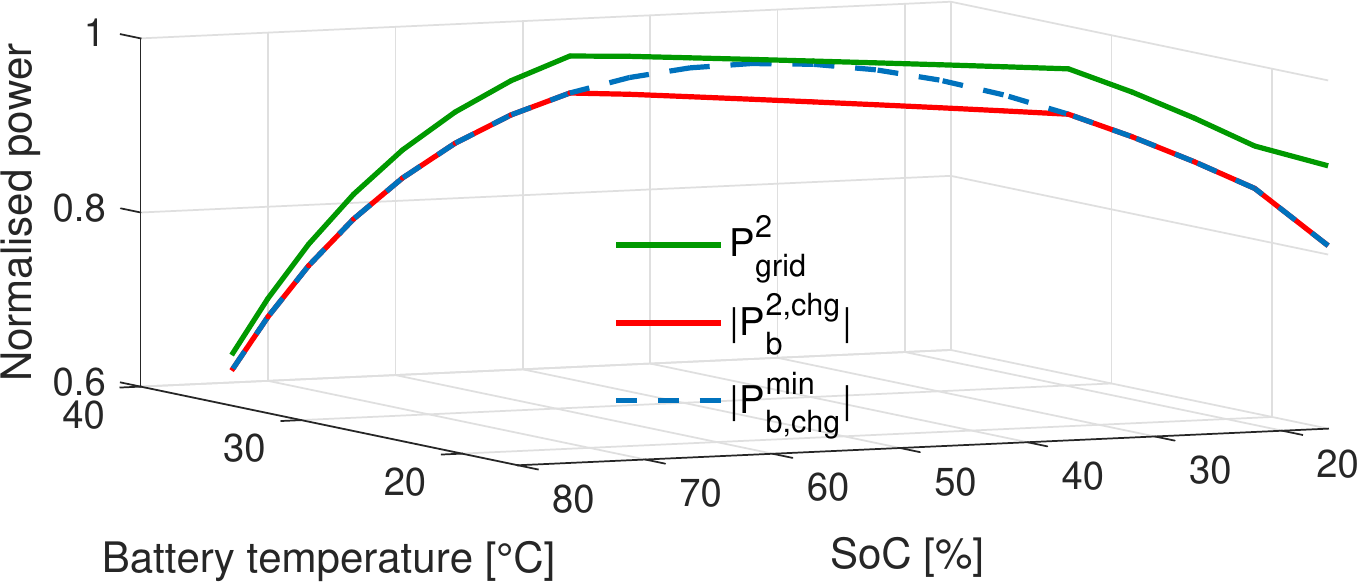}
\label{fig:pbgridtbchg2}
}
\caption{Case 1; optimal trajectories associated with the vehicle's terminal charging mode at $s=\SI{440}{km}$.}
\label{fig:chg2}
\end{figure}

The battery SoC and temperature during terminal charging at $s=\SI{440}{km}$, have similar behaviours as they had during the intermediate charging. In the beginning of charging, initial battery SoC and temperature, respectively, are about $\SI{15}{\%}$ and $\SI{17}{^\circ C}$. HVCH stays on for about a minute from the beginning of charging, and the battery temperature rises up to about $\SI{20}{^\circ C}$ accordingly. Fig.~\ref{fig:pbgridtbchg2} shows a 3D plot including grid power together with the absolute values of battery charging power and its limit for a given combination of SoC and battery temperature. As expected, the charging power availability drops for high SoC values. The charging time is about $\SI{22}{min}$.

\subsection{Case 2: Time Efficient Trip without Active Heating/Cooling}\label{subsec:withoutacthc}
Similar to Section~\ref{subsec:acthc}, the simulation results are summarized into the distance based and time based trajectories. Here, HVCH and HVAC are let to be zero throughout the vehicle's entire trip.

\subsubsection{Optimal Trajectories Vs. Travelled Distance}\label{subsubsection:drv0}
\begin{figure}[t!]
\centering
\subfigure[Road topography together with vehicle speed profile and speed limits.]{
 \includegraphics[width=.92\linewidth]{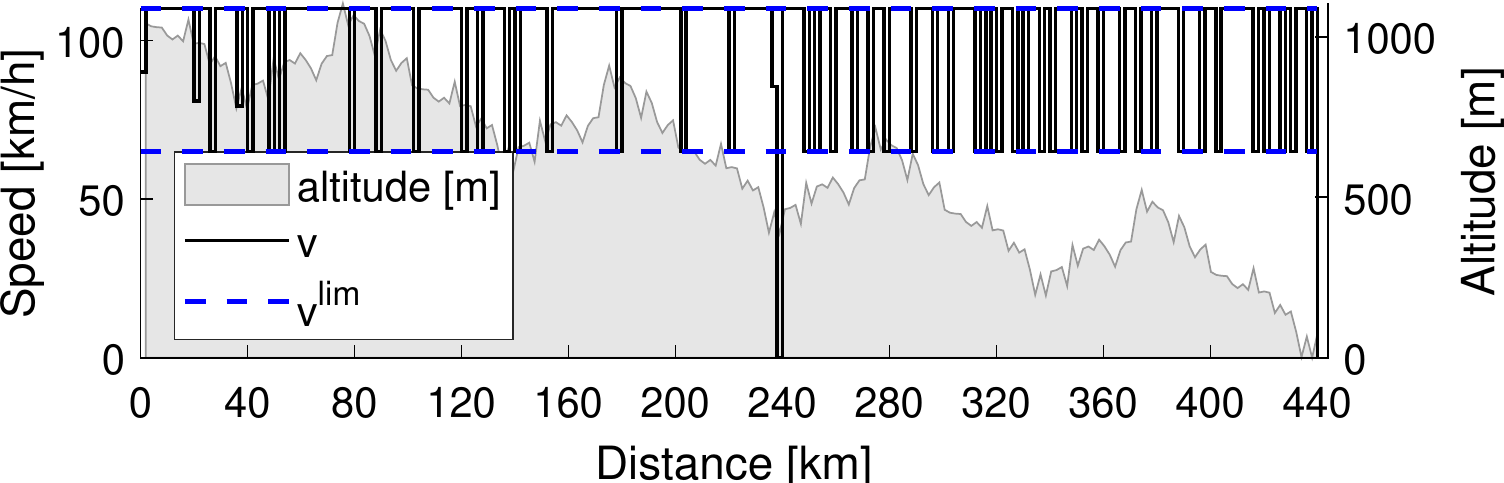}
\label{fig:v0}
}
\subfigure[Battery state of charge trajectory together with its bounds.]{

 \includegraphics[width=.9\linewidth]{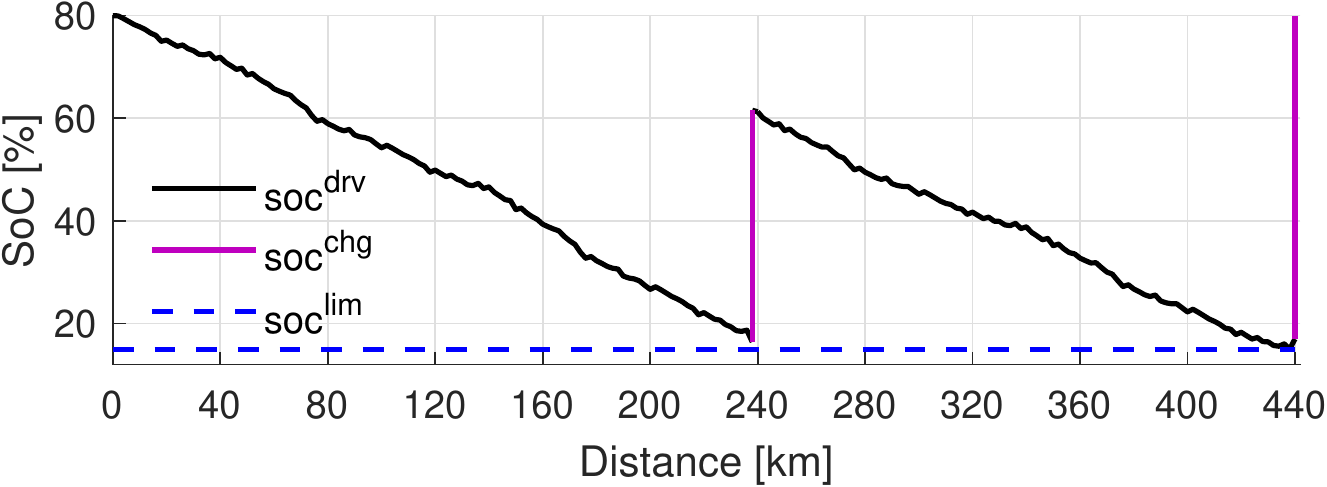}
\label{fig:socdrv0}
}
\subfigure[Battery temperature trajectory together with its upper bound and ambient temperature.]{

 \includegraphics[width=.9\linewidth]{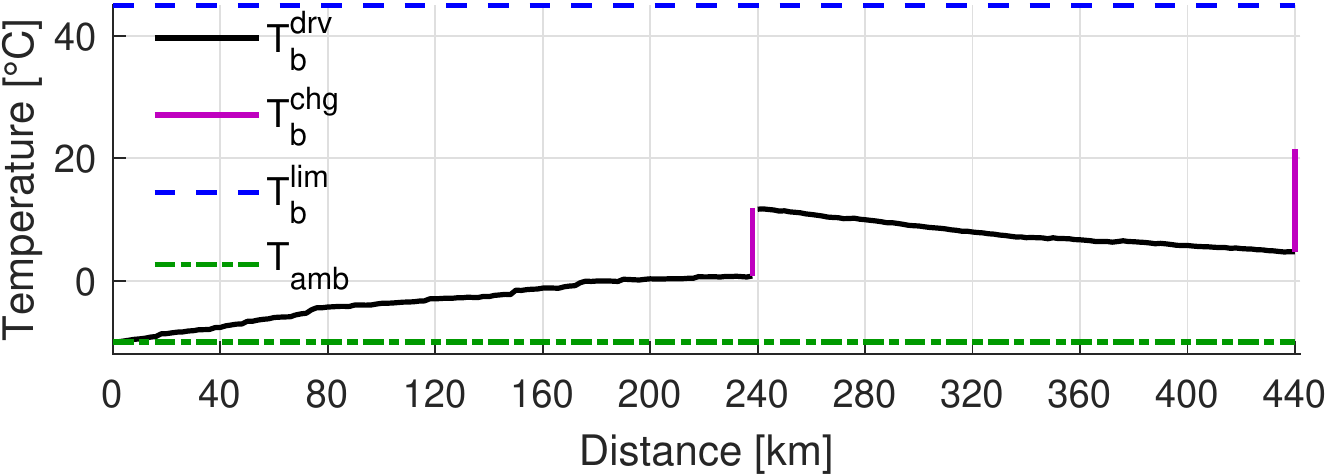}
\label{fig:tbdrv0}
}
% \subfigure[Trajectories of HVCH and HVAC power for battery heating.]{

%  \includegraphics[width=.9\linewidth]{Fig/hvchhvacdrv0.pdf}
% \label{fig:hvchhvacdrv0}
% }
\subfigure[Trajectories of battery power and propulsion power together with battery discharge power limit.]{

 \includegraphics[width=.9\linewidth]{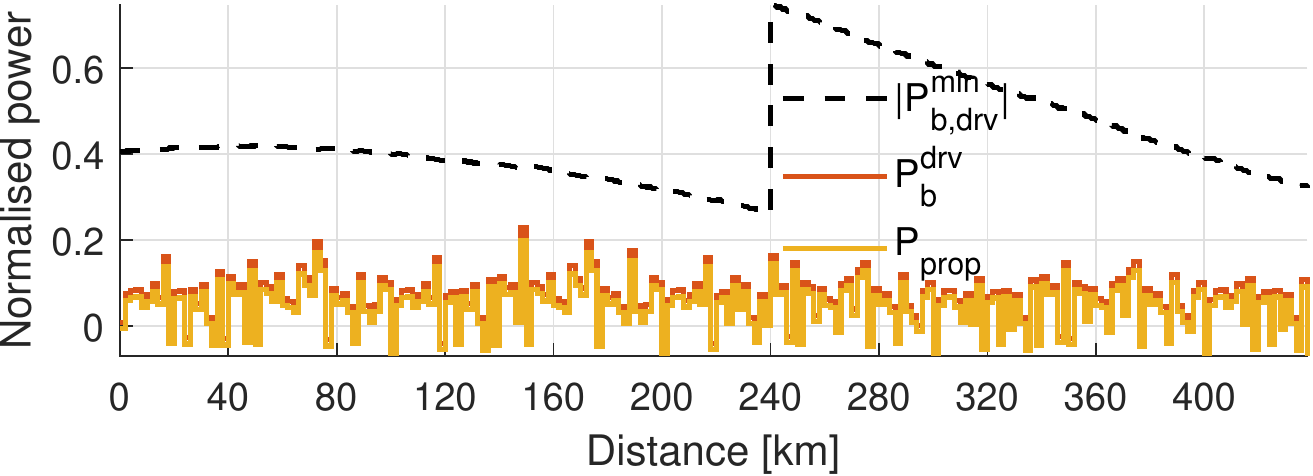}
\label{fig:pbpropdrv0}
}
\caption{Case 2; optimal trajectories Vs. travelled distance in Case 2. The step changes in battery temperature and SoC at $s=\SI{240}{km}$ and $s=\SI{440}{km}$ in (a) and (b), denote the increase in the corresponding variables during charging mode.}
\label{fig:drv0}
\end{figure}

Optimal vehicle speed profile as well as the speed limits and road topography are depicted in Fig.~\ref{fig:v0}. The battery depletion profile, shown in Fig.~\ref{fig:socdrv0}, follows a similar trend as the one in Case 1, since in both cases in addition to the identical simulation parameters and the driving behaviour are similar, i.e. $v_\tx{avg}\approx\SI{100}{km/h}$. The SoC levels at the arrival of the charging stations at $s=\SI{240}{km}$ and $s=\SI{440}{km}$, are respectively about $\SI{17}{\%}$ and $\SI{15}{\%}$, as depicted in Fig.~\ref{fig:socdrv0}. The battery temperature increase is simply due to Joule heat and ED losses, according to Fig.~\ref{fig:tbdrv0}, where at the arrival of the intermediate and terminal charging stations, the battery temperature is $\SI{0}{^\circ C}$ and $\SI{5}{^\circ C}$, respectively. These battery temperature values are lower compared to Case 1, as no active heating is applied in Case 2. In the intermediate and terminal charging stations, the battery is charged to about $\SI{60}{\%}$ and $\SI{80}{\%}$ SoC levels, respectively. The propulsion power together with the battery discharge power and its limit are shown in Fig.~\ref{fig:pbpropdrv0}, where the limit is generally lower compared to the one in Case 1, due to the battery's operation in the lower temperature region.

\subsubsection{Intermediate and Terminal Charging}\label{subsubsection:intchg0}
\begin{figure}[t!]
\centering
\subfigure[Battery state of charge trajectory together with its bounds.]{

 \includegraphics[width=.9\linewidth]{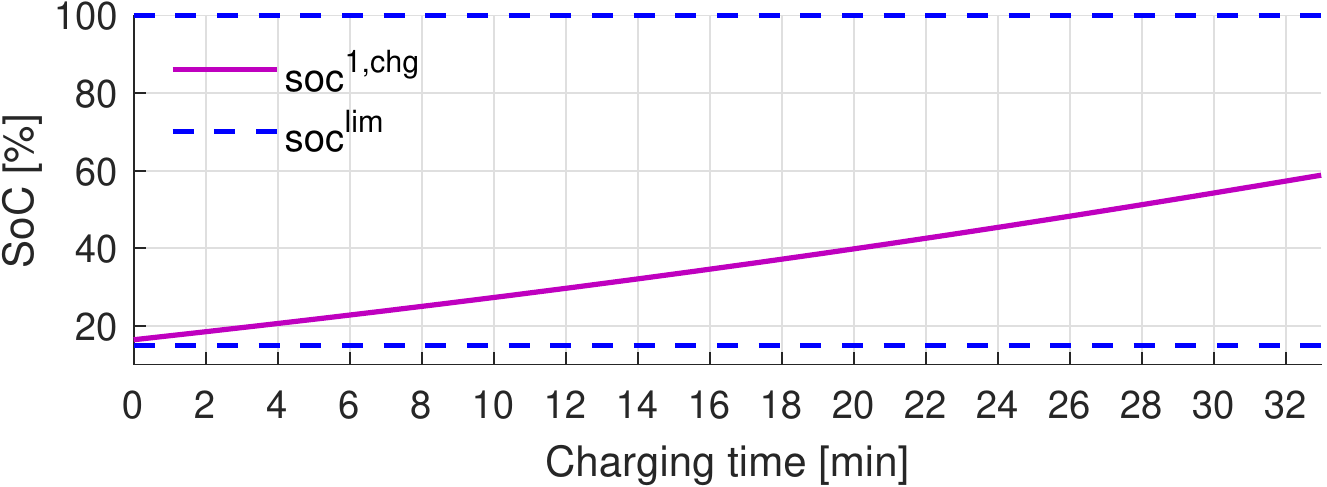}
\label{fig:socchg10}
}
\subfigure[Battery temperature trajectory together with its bounds.]{

 \includegraphics[width=.9\linewidth]{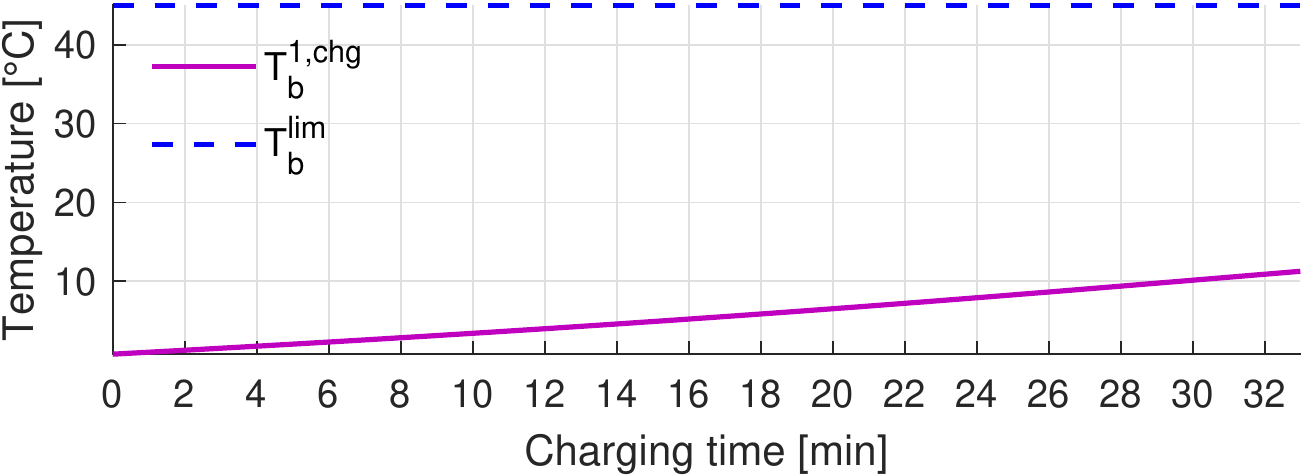}
\label{fig:tbchg10}
}
% \subfigure[Trajectories of HVCH and HVAC power for battery heating.]{

%  \includegraphics[width=.9\linewidth]{Fig/hvchhvacchg10.pdf}
% \label{fig:hvchhvacchg10}
% }
\subfigure[Grid power together with absolute values of battery charging power and its limit, versus battery temperature and SoC.]{

 \includegraphics[width=.9\linewidth]{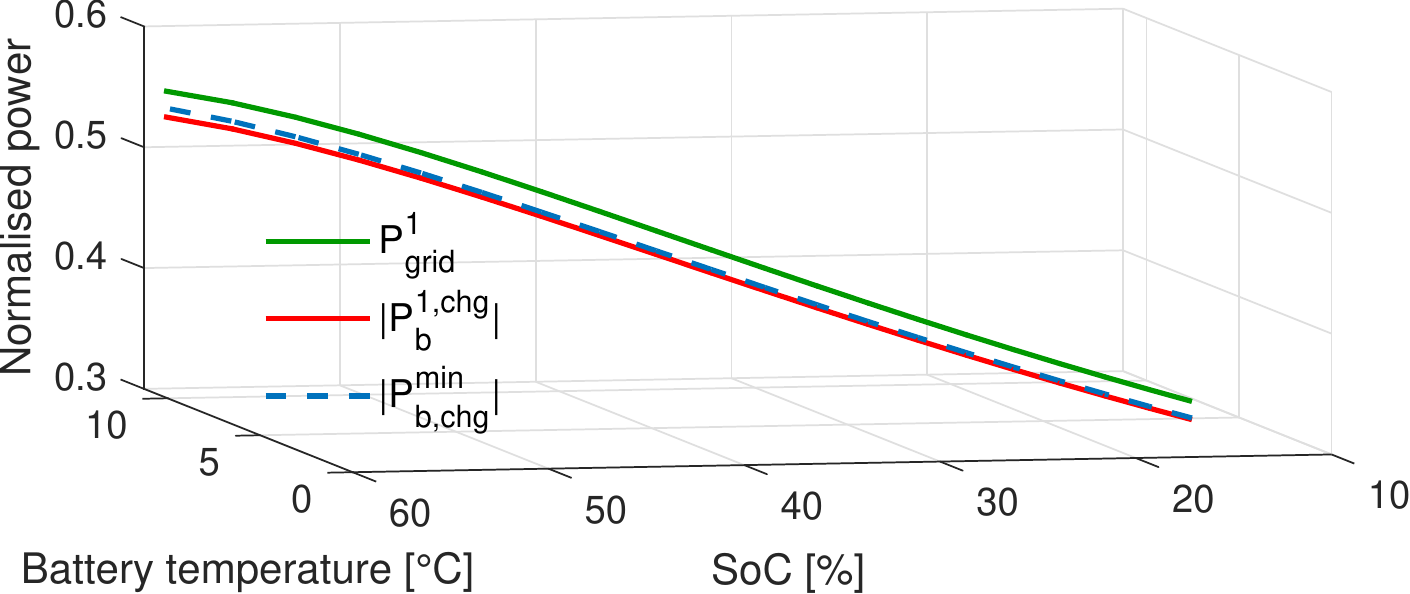}
\label{fig:pbgridtbchg10}
}
\caption{Case 2; optimal trajectories associated with the vehicle's intermediate charging mode at $s=\SI{240}{km}$.}
\label{fig:chg10}
\end{figure}

\begin{figure}[t!]
\centering
\subfigure[Battery state of charge trajectory together with its bounds.]{

 \includegraphics[width=.9\linewidth]{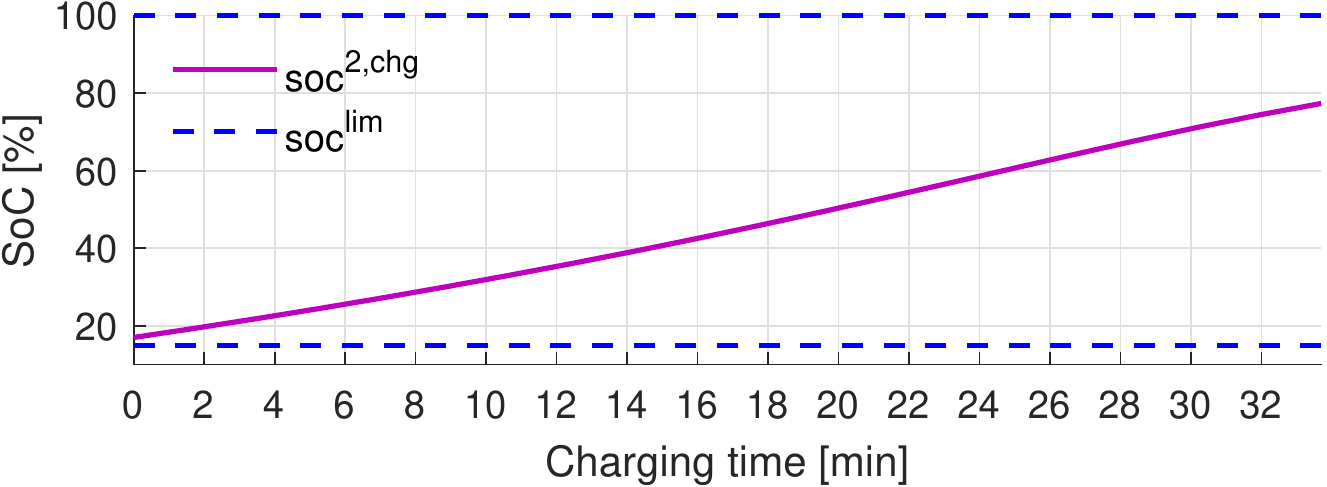}
\label{fig:socchg20}
}
\subfigure[Battery temperature trajectory together with its bounds.]{

 \includegraphics[width=.9\linewidth]{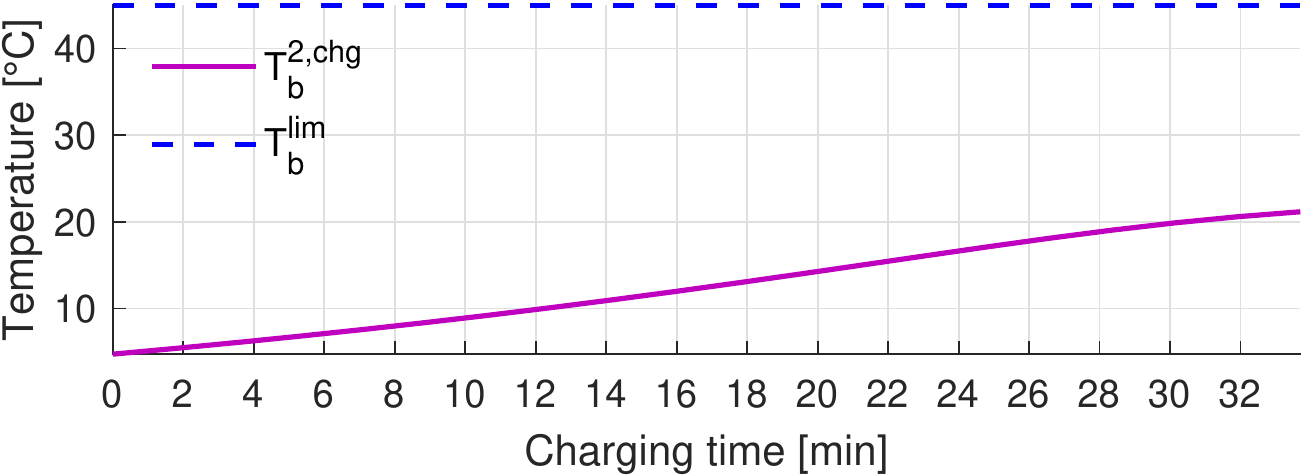}
\label{fig:tbchg20}
}
% \subfigure[Trajectories of HVCH and HVAC power for battery heating.]{

%  \includegraphics[width=.9\linewidth]{Fig/hvchhvacchg20.pdf}
% \label{fig:hvchhvacchg20}
% }
\subfigure[Grid power together with absolute values of battery charging power and its limit, versus battery temperature and SoC.]{

 \includegraphics[width=.9\linewidth]{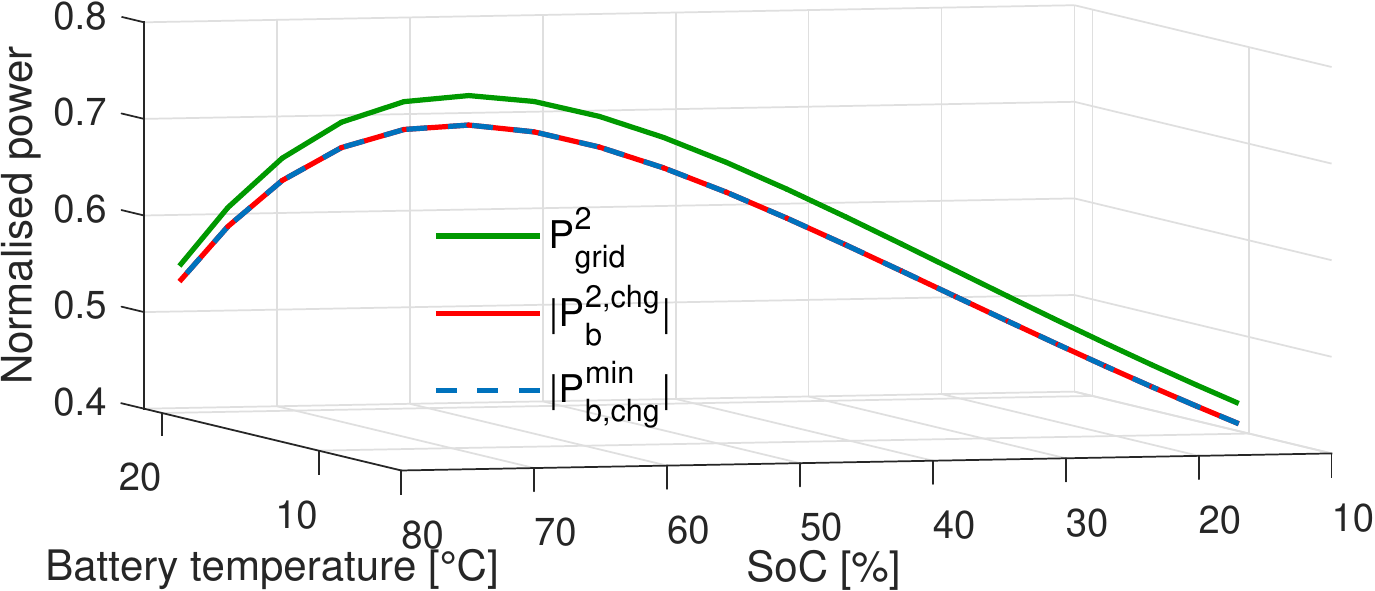}
\label{fig:pbgridtbchg20}
}
\caption{Case 2; optimal trajectories associated with the vehicle's intermediate charging mode at $s=\SI{440}{km}$.}
\label{fig:chg20}
\end{figure}

During both intermediate and terminal charging periods, the battery temperature and SoC increase monotonically, as demonstrated in Fig.~\ref{fig:socchg10} and Fig.~\ref{fig:tbchg10}, and Fig.~\ref{fig:socchg20} and Fig.~\ref{fig:tbchg20}, respectively. Also, the grid power together with the absolute values of battery charging power and its limit versus battery temperature and SoC, are shown in Fig.~\ref{fig:pbgridtbchg10} and Fig.~\ref{fig:pbgridtbchg20}, respectively for the intermediate and terminal charging modes. The charging power availability for Case 2 is lower compared to Case 1, which leads to a higher charging time. According to the results reported in Table \ref{tab:costvst}, total charging time for Case 2 is $\SI{66}{min}$, which is increased by $\SI{44}{\%}$ compared to the Case 1 with optimal battery preconditioning. Instead, the charging cost is simply reduced by $\SI{2}{\%}$.

\section{Conclusion and Future Work}\label{sec:con}
In this paper, optimal BTM, charging, and eco-driving problems are jointly solved for a BEV. To do so, an optimisation problem is formulated, in which the objective includes trip time and charging cost. Later, the problem is reformulated as a hybrid dynamical system, in which the dynamics in each mode, i.e. driving or charging, is modeled with distinct functions, state variables, and control inputs. Furthermore, to reduce computational complexity, the driving dynamics are modeled in a spatial domain, i.e. the decisions are made along the traveled distance. Also, charging dynamics are modeled in a temporal domain, i.e. the decisions are made along the normalized charging time. The actual charging time is treated as a scalar variable, which is optimized simultaneously with the optimal state and control trajectories of both modes. The performance of the
proposed algorithm is evaluated for a vehicle driving on a route, along which two charging possibilities are considered. To study the trade-off between trip time and charging energy cost, the Pareto frontier is derived for different driving scenarios of the vehicles. Such profile is helpful for various car users and vehicle manufacturers to tailor any kind of a trip. According to the results, trip time is reduced by \SI{44}{\%}, in case the optimal battery preconditioning is applied to the vehicle. Low charging time, high charging power availability, and the preservation of the vehicle's potential range are the knock-on effects of the battery preconditioning.

The proposed algorithm for eco-driving and BTM of BEVs can also be extended in several ways, such as: 
\begin{enumerate}
    \item A heat pump can be incorporated in the BTM system to include heating/cooling the battery. In case of the battery cooling, the excess heat from the battery can be transferred to the cabin compartment and/or ambient air. Also, heat pumps are able to transfer the heat from ambient air to the cabin.
    \item It is possible to optimise the charging location to further reduce the trip time.
    \item An online-implementable algorithm can be developed based on the current algorithm that is capable of reacting to potential disturbances, considering model-plant mismatches, and anticipating future events.
\end{enumerate}

\appendices

\section*{Acknowledgment}
The authors would like to acknowledge Mats Bohman, Anand Ganesan, and Ole-Fredrik Dunderberg from Volvo Car Corporation for the support
and helpful discussions during this research. This work is part of a project titled ``Predictive Energy and Thermal management of Electric Vehicles with Connectivity to Infrastructure" funded by Swedish Electromobility Center.

\nocite{lopez17}
\bibliographystyle{IEEEtran}
\bibliography{bibliography_ah}

\end{document}